\documentclass[aps,prl,twocolumn,preprintnumbers,amsmath,amssymb,superscriptaddress]{revtex4-2}
\usepackage{amsmath,graphicx}
\usepackage[utf8]{inputenc}
\usepackage[T1]{fontenc}
\usepackage{xcolor}
\usepackage{textcomp}
\usepackage{bm}

\usepackage{siunitx}
\usepackage{physics}
\usepackage{amsmath}
\usepackage{tikz}
\usepackage{mathdots}
\usepackage{yhmath}
\usepackage{cancel}
\usepackage{color}
\usepackage{array}
\usepackage{multirow}
\usepackage{amssymb}
\usepackage{gensymb}
\usepackage{tabularx}
\usepackage{extarrows}
\usepackage{booktabs}
\usetikzlibrary{fadings}
\usetikzlibrary{patterns}
\usetikzlibrary{shadows.blur}
\usetikzlibrary{shapes}

\usepackage{color}
\definecolor{LinkColor}{rgb}{0.75,0.0,0.2}

\usepackage{hyperref}
\hypersetup{
	pdfauthor={good guys},
	pdftitle={good title},
	colorlinks=true,
	citecolor=LinkColor,
	linkcolor=LinkColor,
	urlcolor=LinkColor,
}

\usepackage{listings}
\definecolor{lightgray}{gray}{1}

\usepackage{theorem}

\def\Eq#1{Eq.~(\ref{#1})}

\usepackage{tikz}

\newcommand{\nc}{\newcommand}
\nc{\braoprket}[3]{\langle#1|#2|#3\rangle}
\nc{\opn}[1]{\operatorname{#1}}
\nc{\avg}[1]{\langle#1\rangle}
\nc{\ketbrasame}[1]{|#1\rangle\!\langle#1|}
\nc{\swap}{\opn{SWAP}}
\nc{\E}{\mathbb{E}}
\nc{\Var}{\opn{Var}}
\nc{\dg}{\dagger}

\usepackage[normalem]{ulem}
\nc{\hknew}[1]{\textcolor{brown}{#1}}

\begin{document}

\title{Bond Additivity and Persistent Geometric Imprints of Entanglement in Quantum Thermalization}

\author{Chun-Yue Zhang}
\affiliation{Beijing National Laboratory for Condensed Matter Physics $\&$ Institute of Physics, Chinese Academy of Sciences, Beijing 100190, China}
\affiliation{University of Chinese Academy of Sciences, Beijing 100049, China}

\author{Shi-Xin Zhang}
\email{shixinzhang@iphy.ac.cn}
\affiliation{Beijing National Laboratory for Condensed Matter Physics $\&$ Institute of Physics, Chinese Academy of Sciences, Beijing 100190, China}

\author{Zi-Xiang Li}
\email{zixiangli@iphy.ac.cn}
\affiliation{Beijing National Laboratory for Condensed Matter Physics $\&$ Institute of Physics, Chinese Academy of Sciences, Beijing 100190, China}
\affiliation{University of Chinese Academy of Sciences, Beijing 100049, China}

\date{\today}

\begin{abstract}
Characterizing the intricate structure of entanglement in quantum many-body systems remains a central challenge, as standard measures often obscure underlying geometric details. In this Letter, we introduce a powerful framework, termed multi-bipartition entanglement tomography, which probes the fine structure of entanglement across an exhaustive ensemble of distinct bipartitions. Our cornerstone is the discovery of a ``bond-additive law'', which reveals that the entanglement entropy can be precisely decomposed into a bulk volume-law baseline plus a geometric correction formed by a sum of local contributions from crossed bonds of varying ranges. This law distills complex entanglement landscapes into a concise set of entanglement bond tensions $\{\omega_j\}$, serving as a quantitative fingerprint of interaction locality. By applying this tomography to Hamiltonian dynamics, random quantum circuits, and Floquet dynamics, we resolve a fundamental distinction between thermalization mechanisms: Hamiltonian thermalized states retain a persistent geometric imprint characterized by a significantly non-zero $\omega_1$, while this structure is completely erased in random quantum circuit and Floquet dynamics. Our work establishes multi-bipartition entanglement tomography as a versatile toolbox for the geometric structure of quantum information in many-body systems.
\end{abstract}

\maketitle

\textit{Introduction.---}Entanglement entropy (EE) has become a central tool for understanding non-equilibrium quantum dynamics, from characterizing thermalization to revealing information scrambling \cite{Calabrese_2004,
RevModPhys.82.277,
doi:10.1126/science.aaf6725,
RevModPhys.90.035007,
PhysRevLett.109.040502,
PhysRevB.95.094206,
PhysRevX.9.021033,
ycdh-z8zf}. In both thermal Hamiltonian systems \cite{PhysRevA.43.2046,
PhysRevE.50.888,
Srednicki_1999,
rigol_thermalization_2008,
Deutsch_2018,
Gogolin_2016,
PhysRevLett.105.250401,
PhysRevLett.111.127205,
PhysRevLett.122.250602,
PhysRevLett.131.060401,
PhysRevX.14.041051,
PRXQuantum.2.010340,
PhysRevB.99.155130,
PhysRevB.105.075117,
PhysRevA.80.053607,
PhysRevA.82.011604,
PhysRevE.79.061103,
PhysRevE.81.036206,
PhysRevE.85.060101,
PhysRevE.90.050101,
PhysRevE.103.052213,
PhysRevE.106.044103,
PhysRevE.110.024135,
PhysRevE.110.044127,
D'Alessio03052016,
mezei_entanglement_2017,
bianchi_linear_2018,
xxlq-d1sw} and generic random quantum circuits (RQCs) \cite{PhysRevLett.98.130502,
PhysRevX.7.031016,
PhysRevX.8.021013,
PhysRevX.8.021014,
PhysRevB.99.174205,
PhysRevB.100.134306,
Skinner2019PRX,
Jian2020PRBmitp,
PhysRevX.11.021040,
PhysRevB.107.L201113,
fisher2023random,
Han2023entanglement,
PhysRevB.110.064323,
PhysRevLett.132.240402,
PhysRevLett.133.140405,
wang2024drivencriticaldynamicsmeasurementinduced,
liu2025noisymonitoredquantumcircuits,
Chen2025subsystem,
10.21468/SciPostPhys.19.5.132}, the EE typically exhibits rapid growth over time and saturates near the Haar measure average value \cite{PhysRevLett.71.1291,
PhysRevE.93.052106,
PhysRevD.100.105010,
PhysRevB.108.245101,
PhysRevLett.133.070402}. Within this prevailing paradigm, the late-time evolved state is regarded as a featureless volume-law ensemble, having effectively discarded memory of its microscopic interaction structure and initial-state configurations \cite{goldstein_long-time_2010,
PhysRevLett.108.080402,
Gogolin_2016,
PhysRevLett.128.060601,
PhysRevLett.131.250401,
PhysRevX.14.041059,
PhysRevX.15.011031,
bwr6-vskn}.

However, a fundamental conceptual tension remains: unlike random circuits, Hamiltonian dynamics are strictly constrained by conservation laws, most notably energy conservation \cite{PhysRevA.101.042126,
PhysRevX.14.031014,
xxlq-d1sw,
mao2025randomunitariesconserveenergy,
cui2025randomunitarieshamiltoniandynamics}. While it is recognized that this constraint prevents the system from exploring the full Hilbert space like a Haar-random unitary, the precise structural consequences for quantum entanglement remain unexplored. Current paradigms often view the Hamiltonian thermalized state as featureless within an energy shell, leaving an important question unanswered: does the locality of physical interactions leave a persistent, discernible fingerprint on the geometry of entanglement for the Hamiltonian thermalized states?

In this Letter, we resolve this mystery by introducing multi-bipartition entanglement tomography---a framework that systematically probes the fine structure of entanglement across a vast ensemble of geometrically distinct bipartitions. Unlike standard approaches that rely on a single fixed cut, e.g., the conventional half-chain EE, our method provides a comprehensive, high-resolution characterization of the entanglement landscape. By parameterizing each bipartition according to its boundary geometry, specifically, the number of ``crossed bonds'' $\{n_j\}$ spanning different ranges (see Fig.~\ref{fig:bipartition_and_thermal_sat_EE}(a)), we transform a single entropy value into a multi-dimensional map of quantum information.

This comprehensive geometric description leads to our central discovery: the EE of Hamiltonian thermalized states remarkably obeys a simple \textit{bond-additive law}. This principle posits that the EE of an arbitrary bipartition can be precisely decomposed into a bulk volume-law baseline and a sum of local contributions from crossed bonds, distilling the complex entanglement landscape into a concise set of entanglement bond tensions $\{\omega_j\}$. These tensions serve as a quantitative fingerprint, capturing both the Hamiltonian’s underlying locality and the fine structure of the quantum entanglement. As a glimpse, Fig.~\ref{fig:bipartition_and_thermal_sat_EE}(b) reveals a highly structured distribution in the saturated EEs of thermal Hamiltonian, manifested as a pronounced linear dependence on the number of nearest-neighbor (NN) crossed bonds ($n_1$), that is entirely absent in RQC and Floquet dynamics. Specifically, we identify a robust hierarchy, $\omega_1 \gg \omega_j$ for $j > 1$, which quantitatively captures the enduring geometric imprint of Hamiltonian locality. Our findings demonstrate that energy conservation compels the system to thermalize coherently along a persistent geometric skeleton defined by the locality of its Hamiltonian. Ultimately, our work establishes a versatile framework for deciphering the geometric origins of entanglement and provides new insights into the paradigm of thermalization in quantum many-body systems.

\begin{figure}[!htbp]
    \centering
    \includegraphics[width=0.9\columnwidth]{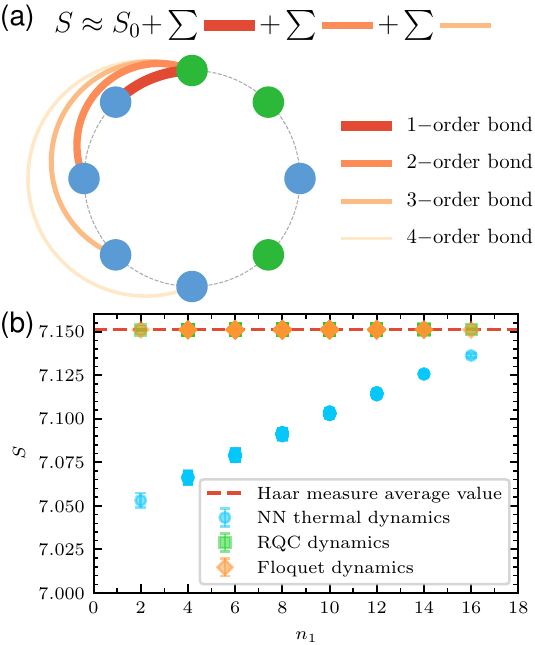}
    \caption{
    (a) Schematic illustration of a bipartition for a chain with periodic boundary condition. The spins represented by the green and blue circles constitute subsystem $\text{A}$ and its complement, respectively. Examples of crossed bonds with different orders connecting to the top spin are shown.
    (b) Saturated EEs of equal-sized bipartitions as a function of the crossed $1$-order bonds number $n_1$ for chain length $L=16$. The NN thermal dynamics governed by $\hat{H}_{\text{NN}}(W=0.5)$ (evolution time $t=1000.0$) shows a structured distribution that remains below the Haar measure average value. In contrast, RQC (evolution depth $2000$) and Floquet (evolution period number $100$) dynamics make saturated EEs perfectly reproduce the Haar measure average. All these results are averaged over $1000$ random samples.
    }
    \label{fig:bipartition_and_thermal_sat_EE}
\end{figure}

\textit{Model and Methods.---}We investigate the dynamics of a one-dimensional spin-$1/2$ chain of length $L$ with periodic boundary conditions. Our primary model is the disordered XXZ Hamiltonian with NN spin interactions:
\begin{equation}
\hat{H}_{\text{NN}}=\sum_{i=1}^{L}\left(\hat{S}_{i}^x\hat{S}_{i+1}^x+\hat{S}_{i}^y\hat{S}_{i+1}^y+J_z\hat{S}_{i}^z\hat{S}_{i+1}^z + h_i\hat{S}_{i}^z\right),
\label{eq:H_NN}
\end{equation}
where $\hat{S}^{\alpha}_i$ are spin-$1/2$ operators on site $i$, $J_z = 0.5$, and random fields $h_i \in [-W, W]$ are drawn uniformly. We set $W=0.5$ to ensure the system resides in the ergodic thermal phase (see Supplemental Material (SM) Sec.~\ref{sec:sm_level_spacing} for phase diagnostics). Thus, $\hat{H}_{\text{NN}}(W=0.5)$ govern a \textit{NN thermal dynamics}. All results are obtained by initializing the system in a product state.

\begin{figure*}[thtbp]
    \centering
    \includegraphics[width=1.0\textwidth]{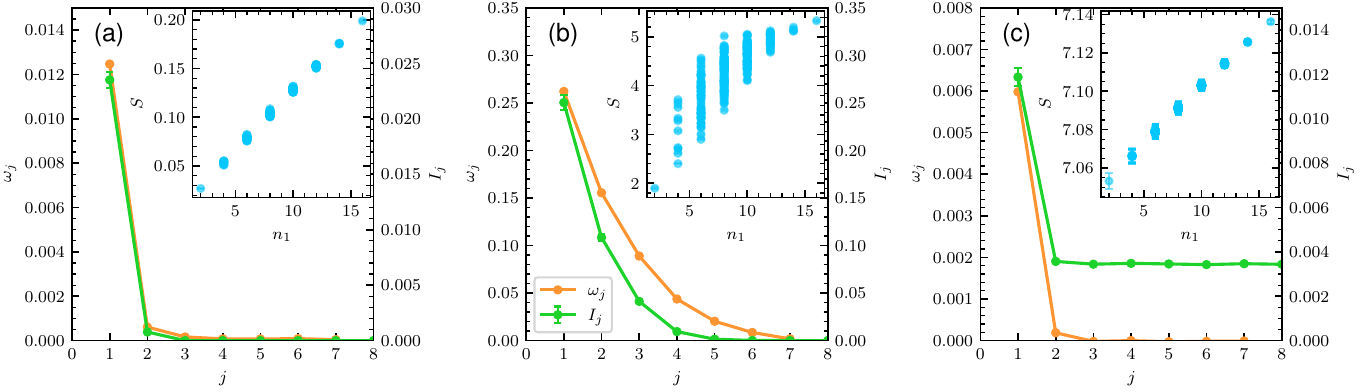} 
    \caption{
    Evolution of the entanglement structure and mutual information in NN thermal dynamics for $L=16$.
    The panels compare the entanglement bond tensions $\left\lbrace \omega_j\right\rbrace$ and the mutual informations $\left\lbrace I_j\right\rbrace$ at three distinct stages: (a) early time ($t=0.1$), (b) intermediate time ($t=2.0$), and (c) late time ($t=1000.0$) at which the entanglement has saturated.
    The entanglement bond tensions $\left\lbrace \omega_j\right\rbrace$ are extracted from linear regression on the EEs of equal-sized bipartitions ($n_0=L/2=8$), whose raw distributions against $n_1$ are shown in the corresponding insets.
    The excellent quality of the fits is confirmed by the high coefficients of determination, $R^2 \approx 0.9993$ (a), $R^2 \approx 0.9959$ (b), and $R^2 \approx 0.9995$ (c).
    The persistence of the hierarchy $\omega_1 \gg \omega_j$ for $j>1$ in panel (c) demonstrates a lasting imprint of locality in the thermalized state.
    All the data of mutual informations and EEs used for fitting are averaged over $1000$ random samples.
    }
    \label{fig:NN_thermal_early_intermediate_late_omega_I}
\end{figure*}

Our primary probe is the von Neumann EE:
\begin{equation}
S=-\operatorname{Tr}\left(\hat{\rho}_\text{A}\log_2\hat{\rho}_\text{A}\right),
\label{eq:EE}
\end{equation}
where $\hat{\rho}_\text{A}$ is the reduced density operator of a subsystem $\text{A}$ which contains $n_0$ spins. By analyzing the EEs for distinct bipartitions of subsystem $\text{A}$ (see SM Sec.~\ref{sec:sm_bipartitions_symmetry_analysis} for details), we uncover the fine structure of entanglement and reveal a persistent geometric imprint of the Hamiltonian's locality throughout the entire evolution. To quantitatively describe this fine structure, we characterize each bipartition not only by its size $n_0$ but also by its detailed geometric properties. Conceptually, any pair of spins in the chain can be connected by a bond. A bond connecting spins separated by a distance $j$ is termed a $j$-order bond. For a given bipartition, a bond is defined to be ``crossed'' if its two endpoint spins reside in the subsystem $\text{A}$ and its complement, respectively (see Fig.~\ref{fig:bipartition_and_thermal_sat_EE}(a)). We denote the number of crossed $j$-order bonds as $n_j$, e.g., $n_1$ represents the number of domain walls for the given bipartition. The set $\left\lbrace n_j\right\rbrace$ provides a geometric label for each bipartition.

\textit{Results.---}The unique nature of Hamiltonian thermalization becomes immediately apparent when we examine the entanglement structure through our geometric lens. As a first illustration, Fig.~\ref{fig:bipartition_and_thermal_sat_EE}(b) plots the saturated EEs for all symmetry-inequivalent bipartitions with equal-sized subsystems ($n_0=L/2$) as a function of the number of crossed NN bonds, $n_1$. The result is striking: for the thermal state evolved under Hamiltonian $\hat{H}_{\text{NN}}(W=0.5)$, the values of EE form a highly structured distribution with a dominant linear dependence on $n_1$. This stands in significant contrast to the dynamics of RQCs and the Floquet system (see SM Sec.~\ref{sec:sm_dynamics_description} for detailed setup), where all the saturated EEs collapse to almost completely the Haar measure average value, irrespective of the boundary geometry $n_1$ of the bipartition. This pronounced dependence on local boundary structure provides a compelling first signature of persistent locality within the thermalized state under Hamiltonian dynamics. However, fully resolving the finer geometric architecture of entanglement requires a systematic analysis with the contributions of longer-range bonds.

To fully dissect this geometric entanglement structure, we introduce the framework of \textit{multi-bipartition entanglement tomography}. This framework leverages the EEs under a comprehensive set of bipartitions as a high-resolution probe to characterize the system's entanglement landscape. By parameterizing each bipartition with its geometric labels $\{n_j\}$, we can search for underlying quantitative principles. This tomographic approach determines a set of entanglement bond tensions, $\{\omega_j\}$, that optimally reconstruct the EE for a fixed subsystem size $n_0$ via the following linear ansatz:
\begin{equation}
S_{\text{fitted}}=S_0+\sum_{j=1}^{L/2-1}\omega_jn_j.
\label{eq:fitting_model}
\end{equation}
Here, the entanglement bond tensions $\{\omega_j\}$ are determined by a least-squares fit to the numerical data. The sum is truncated at $j=L/2-1$ because the set of variables $\left\lbrace n_1,n_2,\cdots,n_{L/2}\right\rbrace$ is not linearly independent due to $\sum_{j=1}^{L/2}n_j=n_0\left(L-n_0\right)$. As exemplified for the NN thermal dynamics in Fig.~\ref{fig:NN_thermal_early_intermediate_late_omega_I}, this ansatz describes the data with extraordinary precision, yielding coefficients of determination $R^2$ consistently near unity. This remarkable agreement reveals a fundamental organizing principle, which we term the \textit{bond-additive law}. Physically, this law interprets the EE as an energy-like cost function where the extensive term $S_0$ provides the volume-law baseline, and the bond-additive terms $\sum \omega_j n_j$ represent the surface tension or geometrical penalty associated with the bipartition boundary. The validity of this law implies that, even within the volume-law entanglement, the system retains a fine-grained structure where cutting a specific $j$-order bond incurs a quantized entanglement penalty $\omega_j$.

\begin{figure*}[thtbp]
    \centering
    \includegraphics[width=1.0\textwidth]{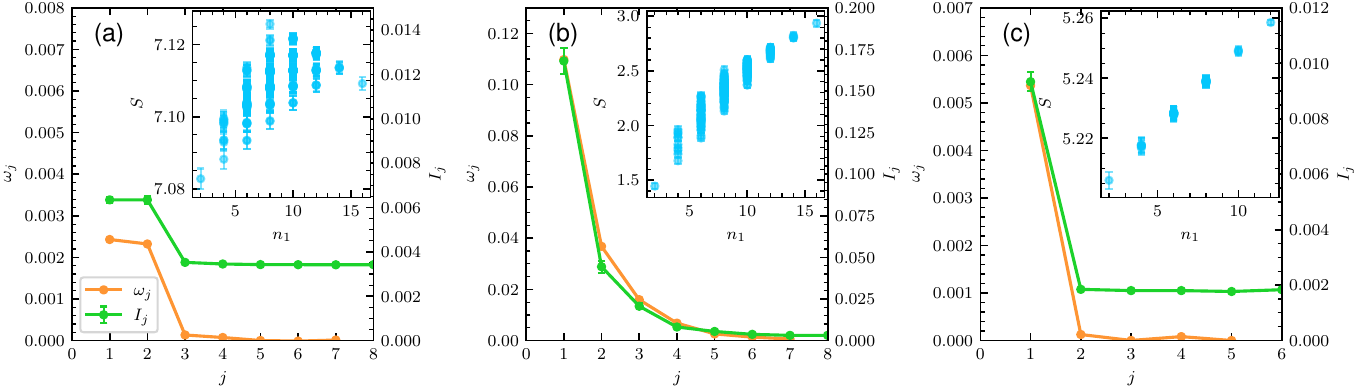} 
    \caption{
    Entanglement bond tensions $\left\lbrace \omega_j\right\rbrace$ with $n_0=L/2$ and mutual informations $\left\lbrace I_j\right\rbrace$ at saturation of other Hamiltonian dynamics. (a) NNN thermal dynamics ($L=16$, $t=1000.0$, $R^2\approx0.9986$). (b) MBL dynamics ($L=16$, $t=10^{12}$, $R^2\approx0.9984$). (c) mixed-field dynamics ($L=12$, $t=1000.0$, $R^2\approx0.9990$). All these results are averaged over $1000$ random samples.
    }
    \label{fig:NNN_thermal_MBL_MF_late_omega_I}
\end{figure*}

The evolution of these entanglement bond tensions, $\left\lbrace\omega_j\right\rbrace$, provides a detailed, dynamical picture of how entanglement geometry is shaped by the local, energy-conserving Hamiltonian. We now apply this analysis to the NN thermal dynamics, with the results shown in Fig.~\ref{fig:NN_thermal_early_intermediate_late_omega_I}. At an early time ($t=0.1$, Fig.~\ref{fig:NN_thermal_early_intermediate_late_omega_I}(a)), the dynamics are dominated by the Hamiltonian's locality. Entanglement is primarily generated between NN spins, which is perfectly captured by a sharp hierarchy in the entanglement bond tensions: $\omega_1 \gg \omega_j$ for $j>1$. As the system evolves to an intermediate time ($t=2.0$, Fig.~\ref{fig:NN_thermal_early_intermediate_late_omega_I}(b)), entanglement propagates through the chain. This is reflected in the significant growth of longer-range entanglement bond tensions such as $\omega_2$, $\omega_3$, $\cdots$, leading to a more scattered EE distribution as seen in the inset. Most strikingly, at the late time ($t=1000.0$, Fig.~\ref{fig:NN_thermal_early_intermediate_late_omega_I}(c)) when the EE has saturated, this hierarchy persists. While the absolute values of $\left\lbrace\omega_j\right\rbrace$ have decreased as the EE becomes dominated by a large volume-law term $S_0$, the shortest-range contribution $\omega_1$ remains significantly larger than the others. This central result reveals that the state evolved by a thermal Hamiltonian for a long enough time is not structurally random but retains a strong geometric imprint of the Hamiltonian's locality. This persistence is a direct consequence of energy conservation of local Hamiltonians, a fundamental constraint absent in RQCs and Floquet systems, where the entanglement geometry is completely washed out upon saturation, leading to a featureless state where all entanglement bond tensions $\left\lbrace\omega_j\right\rbrace$ approach zero.

To understand the microscopic origin of this persistent pattern, we track the evolution of $\{\omega_j\}$ alongside the two-point mutual information $I_j=S_{\{1\}}+S_{\{1+j\}}-S_{\{1,1+j\}}$, where $S_{\{1\}}$, $S_{\{1+j\}}$ and $S_{\{1,1+j\}}$ are respectively the entropies of the first spin, the $(1+j)$-th spin and their union. $I_j$ quantifies the correlation between two spins separated by a distance $j$. As shown in Fig.~\ref{fig:NN_thermal_early_intermediate_late_omega_I}, the evolution of $\{I_j\}$ mirrors that of $\{\omega_j\}$. Initially, correlations build up locally, with $I_1$ dominating. Subsequently, correlations propagate, leading to the growth of longer-range $I_j$. Most importantly, even after the system has saturated to a thermal state at late time, a distinct hierarchy $I_1>I_j$ for $j>1$ persists. This behavior provides a clear, microscopic picture that complements our findings from the multi-bipartition entanglement tomography. The enduring dominance of the NN correlation $I_1$ is a direct signature of the Hamiltonian's underlying locality. This persistent hierarchy is a consequence of energy conservation, which constrains the system's exploration of Hilbert space and prevents the formation of a featureless Haar-random state.  However, it is important to distinguish the roles between $\{I_j\}$ and $\{\omega_j\}$. While $I_j$ detects the presence of local correlations, it is a purely local measure. The bond-additive law reveals a far more non-trivial many-body phenomenon as these local correlations do not scramble into a complex, non-linear web, but instead add up coherently to determine the global EE of macroscopic subsystems.

To explore how the bond-additive law holds and how the entanglement geometry manifests for other Hamiltonian dynamical systems, we investigated the following three other scenarios (all of their detailed setups are in SM Sec.~\ref{sec:sm_dynamics_description}). The values of $R^2$ that are consistently very close to unity indicate that the bond-additive law also holds in these cases.

Our analysis has established that the persistent $\left\lbrace\omega_j\right\rbrace$ hierarchy is a direct fingerprint of the Hamiltonian's interaction range. A natural and powerful test of this hypothesis is to examine a system with a modified locality. To this end, we perform the same tomography on a \textit{next-nearest-neighbor (NNN) thermal dynamics} which is governed by $\hat{H}_{\text{NNN}}$ that is extended to include NNN spin interactions. Fig.~\ref{fig:NNN_thermal_MBL_MF_late_omega_I}(a) shows the results of the saturated state under $\hat{H}_{\text{NNN}}$, which perfectly align with our hypothesis. The addition of NNN couplings significantly boosts the second entanglement bond tension $\omega_2$, elevating it to a value comparable to $\omega_1$. This change in the entanglement geometry is mirrored in the two-point correlations, where the mutual information $I_2$ also becomes comparable to $I_1$. This result provides compelling validation for our multi-bipartition entanglement tomography: the entanglement bond tensions $\left\lbrace\omega_j\right\rbrace$ are not just abstract fitting parameters but serve as a quantitative and predictive measure of the Hamiltonian's underlying interaction geometry.

To explore a non-ergodic phase, we apply our method to many-body localized (MBL) \cite{PhysRevB.77.064426,
PhysRevB.82.174411,
PhysRevLett.109.017202,
PhysRevLett.110.067204,
altman2015universal,
doi:10.1126/science.aaa7432,
PhysRevLett.121.206601,
doi:10.1126/science.aau0818,
FAN2017707,
https://doi.org/10.1002/andp.201600318,
LIU20253991,
PhysRevB.95.054201,
PhysRevB.95.165136,
PhysRevX.5.031032,
Sierant_2025} dynamics, which is governed by $\hat{H}_{\text{NN}}(W=5.0)$. The saturated state exhibits a profoundly different entanglement structure, as shown in Fig.~\ref{fig:NNN_thermal_MBL_MF_late_omega_I}(b). The inset shows an obviously small and highly dispersive EE distribution, a hallmark of non-ergodicity. The structure of our quantitative probes further reveals the specific details that distinguish MBL from thermalization. The leading entanglement bond tension $\omega_1$ and mutual information $I_1$, which capture the shortest-range entanglement and correlation, are more than one order of magnitude larger than those in the thermal case. Besides, the long tail of $\omega_j$ reflects the phenomenological l-bit picture of MBL \cite{PhysRevLett.111.127201,
PhysRevB.90.174202,
PhysRevB.91.085425,
nandkishore2015many,
ROS2015420,
PhysRevLett.116.010404,
https://doi.org/10.1002/andp.201600278,
RevModPhys.91.021001,
PhysRevLett.133.126502}, where entanglement is generated by effective interactions that decay exponentially with distance but remain non-zero at long ranges. 

Finally, we demonstrate the universality of these findings by extending our analysis to dynamics lacking $U(1)$ symmetry. We first consider a disordered mixed-field Hamiltonian, $\hat{H}_{\text{MF}}$, which breaks particle-number conservation while remaining strictly energy-conserving. Evolving initial product states that span multiple charge sectors, we find that the late-time entanglement structure [Fig.~\ref{fig:NNN_thermal_MBL_MF_late_omega_I}(c)] exhibits a hierarchy of $\{\omega_j\}$ and $\{I_j\}$ qualitatively identical to the $U(1)$-symmetric cases. This confirms that the geometric memory of entanglement is a robust consequence of energy conservation, persisting even when the Hilbert space exploration is no longer restricted to a single particle-number sector. A similar robustness is observed for the original $\hat{H}_{\text{NN}}$ model when initialized with sector-unrestricted product states (see SM Sec.~\ref{sec:sm_fitting_details}). These results establish the geometric imprint of locality as a universal signature of Hamiltonian thermalization, independent of additional symmetry constraints.

\textit{Discussion and Conclusion.---}In this Letter, we introduce multi-bipartition entanglement tomography, a powerful framework for characterizing the fine geometric structure of entanglement in quantum many-body systems. By applying this framework to Hamiltonian dynamics initialized from product states, we discovered an extraordinarily precise organizing principle: the bond-additive law. This law distills the complex entanglement landscape into a concise set of entanglement bond tensions, $\{\omega_j\}$, which serve as a quantitative fingerprint of the system's entanglement geometry.

Our framework reveals a fundamental distinction in thermalization rooted strictly in energy conservation. While the prevailing paradigm often regards late-time thermalized states as featureless volume, our results demonstrate that Hamiltonian thermalization is intrinsically structured. The persistent hierarchy of entanglement bond tensions ($\omega_1\gg \omega_j$) demonstrates that energy conservation, as a rigid constraint, prevents the system from erasing its geometric memory. This contrasts with the featureless thermal states produced by RQC and Floquet dynamics. These findings have far-reaching implications for the fundamental theory of quantum thermalization, establishing that the Hamiltonian dynamics retains a fine-grained information structure.

Furthermore, multi-bipartition entanglement tomography provides a powerful lens through which to characterize quantum states.
While recent studies have increasingly underscored the importance of multi-bipartition perspectives in understanding many-body systems~\cite{PhysRevB.101.195134,
PhysRevB.101.224202,
PhysRevB.102.134203,
PhysRevB.103.174309,
kolisnyk2025tensorcrossinterpolationpurities,
zhang2025entanglementgrowthentangledstates,
driebschoen2025bundlingbipartiteentanglement}, our work establishes the direct and quantitative bridge between the global geometry of entanglement and the local interaction patterns of the Hamiltonian. By demonstrating this correspondence, we provide a universal and quantitative framework to distinguish and classify diverse quantum phases and their non-equilibrium dynamics through the geometric structure of quantum entanglement.

Looking ahead, the versatility of multi-bipartition tomography opens several exciting avenues.  Its extension to higher dimensions provides a promising route for isolating universal subleading corrections to the entanglement area law, such as the topological EE in topologically ordered states~\cite{kitaev2006prl,levin2006prl} or the corner coefficients that characterize critical systems and conformal field theories~\cite{1dEE_scaling2,Fradkin2006PRL}. Additionally, our approach offers a strategy for addressing the inverse problem: reconstructing underlying Hamiltonian properties from the spectrum of entanglement bond tensions $\{\omega_j\}$. Experimentally, our framework is readily accessible, as the requisite EEs can be obtained via randomized measurements~\cite{doi:10.1126/science.aau4963} or classical shadow tomography~\cite{huang_predicting_2020,
PhysRevA.106.012441,
PhysRevLett.132.220802,
PhysRevLett.133.060802,
92ky-ln8f,
PhysRevB.111.054306} with no experimental overhead compared to conventional half-chain EE measurements. Our work thus establishes entanglement tomography not just as a conceptual framework, but as a practical and powerful new lens for probing the geometric structure of quantum information in experiments.

\textit{Acknowledgement.---} C.Y.Z. and Z.X.L. are supported by the National Natural Science Foundation of China under Grant Nos. 12347107 and 12474146, and Beijing Natural Science Foundation under Grant No. JR25007. S.X.Z is supported by Quantum Science and Technology-National Science and Technology Major Project (No. 2024ZD0301700) and
the National Natural Science Foundation of China (No. 12574546).

\bibliographystyle{apsreve}
\let\oldaddcontentsline\addcontentsline
\renewcommand{\addcontentsline}[3]{}
\bibliography{ref}
\let\addcontentsline\oldaddcontentsline
\onecolumngrid

\clearpage
\newpage
\widetext

\begin{center}
\textbf{\large Supplemental Material for ``Bond Additivity and Persistent Geometric Imprints of Entanglement in Quantum Thermalization''}
\end{center}

\date{\today}
\maketitle

\renewcommand{\thefigure}{S\arabic{figure}}
\setcounter{figure}{0}
\renewcommand{\theequation}{S\arabic{equation}}
\setcounter{equation}{0}
\renewcommand{\thesection}{\Roman{section}}
\setcounter{section}{0}
\setcounter{secnumdepth}{4}

\addtocontents{toc}{\protect\setcounter{tocdepth}{0}}
{
\tableofcontents
}

\section{Description of Dynamical Protocols}\label{sec:sm_dynamics_description}

We examine distinct dynamical protocols on a one-dimensional spin-$1/2$ chain of length $L$ with periodic boundary condition (PBC). For protocols that conserve $U(1)$ particle number, the initial state $|\psi_0\rangle$ is a computational basis state sampled uniformly from the zero-magnetization (half-filling) sector, $\sum_{i=1}^{L}\hat{S}_{i}^z=0$. We consider five particle-number-conserving protocols defined as follows:

\textit{Nearest-Neighbor (NN) Thermal Dynamics}: This is the primary dynamics we studied, which is governed by a time-independent thermal Hamiltonian:
\begin{equation}
\hat{H}_{\text{NN}}=\sum_{i=1}^{L}\left(\hat{S}_{i}^x\hat{S}_{i+1}^x+\hat{S}_{i}^y\hat{S}_{i+1}^y+J_z\hat{S}_{i}^z\hat{S}_{i+1}^z + h_i\hat{S}_{i}^z\right),
\label{Seq:H_NN}
\end{equation}
where we set the anisotropy to $J_z=0.5$. The on-site random fields $h_i$ are drawn independently and uniformly from the interval $[-W, W]$ and $W=0.5$, thus this Hamiltonian is denoted by $\hat{H}_{\text{NN}}(W=0.5)$.

\textit{Next-Nearest-Neighbor (NNN) Thermal Dynamics}: This dynamics is designed to explore the effect of longer-range interactions. The NNN couplings are introduced:
\begin{equation}
\begin{aligned}
\hat{H}_{\text{NNN}}=\hat{H}_{\text{NN}}(W=0.5)+\gamma\sum_{i=1}^{L}\left(\hat{S}_{i}^x\hat{S}_{i+2}^x+\hat{S}_{i}^y\hat{S}_{i+2}^y+J_z\hat{S}_{i}^z\hat{S}_{i+2}^z\right),
\end{aligned}
\label{Seq:H_NNN}
\end{equation}
where the NNN interaction strength is set to $\gamma=24/25$. This Hamiltonian remains in a thermal phase.

\textit{Many-Body Localized (MBL) Dynamics}: It is realized using the same Hamiltonian as in Eq.~\eqref{Seq:H_NN}, but with a strong disorder strength $W=5.0$ to drive the system into the MBL phase. Its Hamiltonian is denoted by $\hat{H}_{\text{NN}}(W=5.0)$.

\textit{Random Quantum Circuit (RQC) Dynamics}: The circuit is generated by sequentially applying two-qubit unitary gates. In each step, a single gate $\hat{U}_{i,i+1}$ is applied to a randomly chosen adjacent pair of spins $(i, i+1)$ from all $L$ pairs with uniform probability. The unitary is given by:
\begin{equation}
\mathrm{e}^{-\mathrm{i}\pi\left(\hat{S}_{i}^x\hat{S}_{i+1}^x+\hat{S}_{i}^y\hat{S}_{i+1}^y\right)/2}\mathrm{e}^{-\mathrm{i}\pi\hat{S}_{i}^z\hat{S}_{i+1}^z},
\label{Seq:U_RQC}
\end{equation}

\textit{Floquet Dynamics}: The Floquet dynamics is defined by a periodic drive with the Floquet operator:
\begin{equation}
\hat{F}=\mathrm{e}^{-\mathrm{i}T_0\hat{H}_0}\mathrm{e}^{-\mathrm{i}T_1\hat{H}_1},
\label{Seq:Floquet_operator}
\end{equation}
where the two parts of the drive are given by
\begin{equation}
\hat{H}_0=\sum_{i=1}^{L}\left(\hat{S}_{i}^z\hat{S}_{i+1}^z+h_i\hat{S}_{i}^z\right)
\label{Seq:H_0}
\end{equation}
and
\begin{equation}
\hat{H}_1=\sum_{i=1}^{L}\left(\hat{S}_{i}^x\hat{S}_{i+1}^x+\hat{S}_{i}^y\hat{S}_{i+1}^y\right).
\label{Seq:H_1}
\end{equation}
The parameters are set to $T_0=1.0$, $T_1=2.5$, and the disorder strength in $\hat{H}_0$ is $W=5.0$. This setup is adapted from Ref.~\cite{PhysRevLett.114.140401} but implemented with PBC. This protocol is also thermalizing but does not conserve energy due to time-dependence.

Beyond particle number conserved dynamics, we examine two protocols where $U(1)$ symmetry is absent. In these cases, the initial state is a random product state unrestricted by particle-number conservation, allowing the dynamics to explore the full many-body Hilbert space:
\begin{equation}
\left|\psi_0\right\rangle=\prod_{i=1}^{L}\otimes\left(\cos\frac{\theta_i}2\left|\uparrow\right\rangle+\mathrm{e}^{\mathrm{i}\phi_i}\sin\frac{\theta_i}2\left|\downarrow\right\rangle\right),
\label{random_product_state}
\end{equation}
where random parameters $\left\lbrace \theta_i\right\rbrace$ and $\left\lbrace \phi_i\right\rbrace$ cause each spin to independently distribute uniformly on the surface of the Bloch sphere.

We first consider evolution under the thermal Hamiltonian $\hat{H}_{\text{NN}}$ with $W=0.5$ [Eq.~\eqref{Seq:H_NN}]. Unlike configurations restricted to the half-filling sector, these random product states are superpositions spanning multiple magnetization sectors, thereby allowing the dynamics to explore a significantly larger manifold of the many-body Hilbert space.

For the second protocol starting from random product states without $U(1)$ symmetry, we introduce a mixed-field Hamiltonian. This model breaks the conservation of total magnetization by adding a disordered transverse field term to the NN thermal Hamiltonian:
\begin{equation}
\begin{aligned}
\hat{H}_{\text{MF}}=\hat{H}_{\text{NN}}(W=0.5)+\sum_{i=1}^{L}g_i\hat{S}_{i}^x,
\end{aligned}
\label{Seq:H_MF}
\end{equation}
where $\left\lbrace g_j\right\rbrace$ follows the same distribution as $\left\lbrace h_j\right\rbrace$ in $\hat{H}_{\text{NN}}(W=0.5)$ and is independent of $\left\lbrace h_j\right\rbrace$. The corresponding dynamics is called \textit{mixed-field dynamics}.

\begin{figure}[!htbp]
    \centering
    \includegraphics[width=1.0\columnwidth]{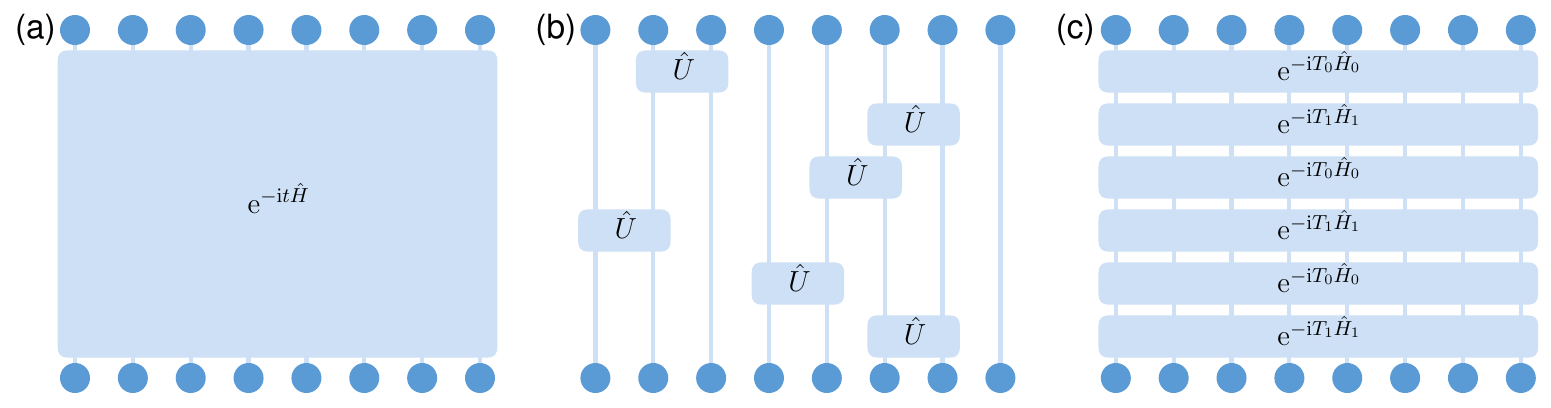}
    \caption{
    Schematic illustrations of Hamiltonian dynamics (where $\hat{H}_{\text{NN}}(W=0.5)$, $\hat{H}_{\text{NN}}(W=5.0)$ $\hat{H}_{\text{NNN}}$ and $\hat{H}_{\text{MF}}$ are all represented by $\hat{H}$) (a), RQC dynamics (b), and Floquet dynamics (c). Blue circles represent spins.
    }
    \label{fig:dynamics_schematics}
\end{figure}

To provide a clear visual representation of the different dynamical protocols, we present their schematic diagrams in Fig.~\ref{fig:dynamics_schematics}. These diagrams illustrate the distinct structures of (a) time-independent Hamiltonian evolution, (b) the sequential gate application in RQCs, and (c) the periodic driving of the Floquet protocol.

\begin{figure}[!htbp]
    \centering
    \includegraphics[width=1.0\columnwidth]{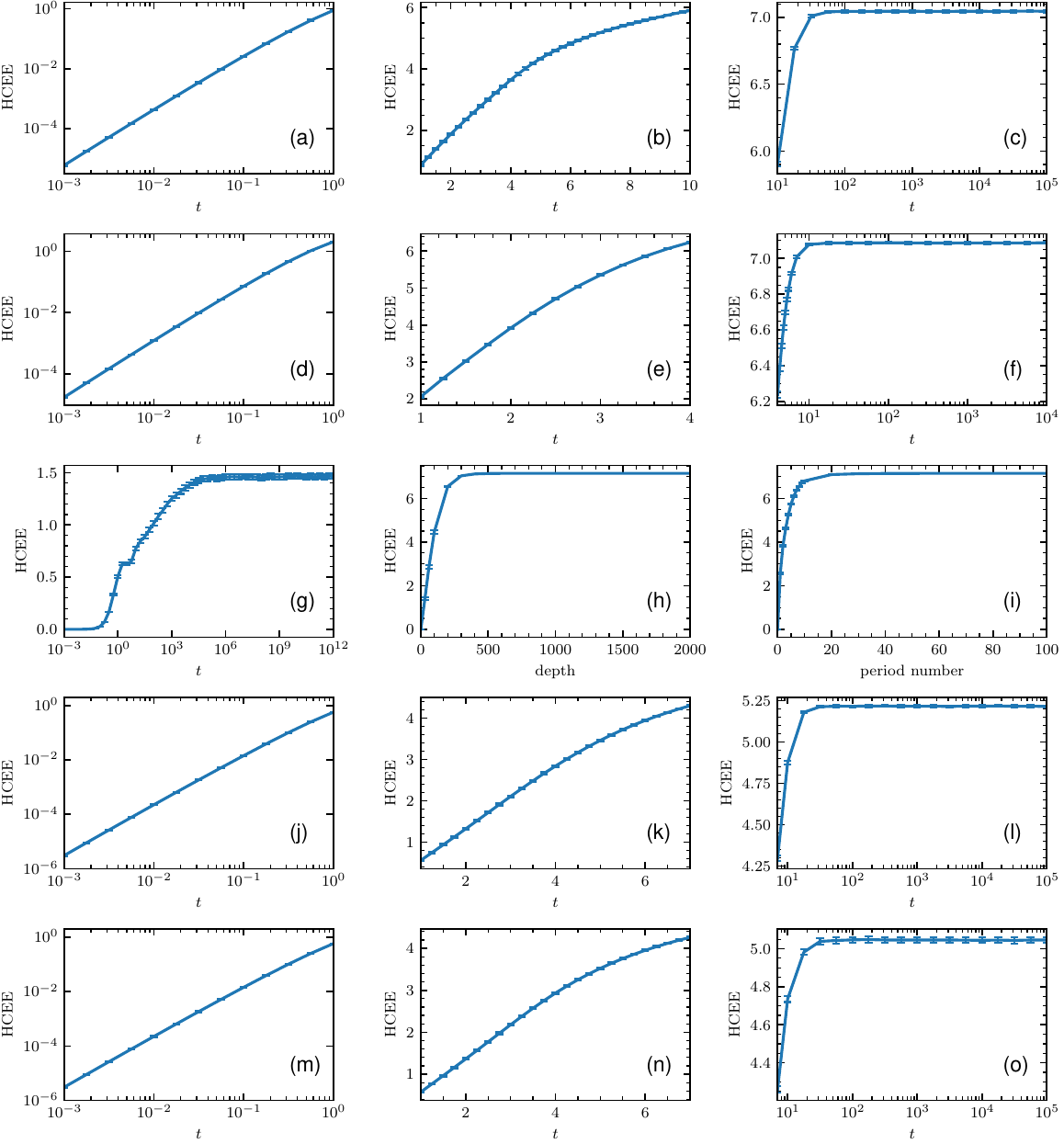}
    \caption{
    Evolution of the HCEE for the various dynamics studied.
    The top row (a-c) displays the HCEE evolution for the NN thermal dynamics, showing the (a) initial linear growth, (b) subsequent slowing, and (c) final saturation.
    Similarly, the second row (d-f), the fourth row (j-l) and the fifth row (m-o) respectively show the evolution for the NNN thermal dynamics, the mixed-field dynamics and the dynamics governed by $\hat{H}_{\mathrm{NN}}(W=0.5)$ starting from random product states across the same three regimes.
    The third row presents the complete HCEE evolution for (g) MBL dynamics, (h) RQC dynamics and (i) Floquet dynamics.
    For the Hamiltonian dynamics (a-g) and (j-o), the horizontal axis is evolution time $t$; for the RQC (h) and Floquet (i) dynamics, it represents the circuit depth and the number of evolution periods, respectively.
    The simulations for mixed-field dynamics and the dynamics governed by $\hat{H}_{\mathrm{NN}}(W=0.5)$ starting from random product state are performed on a chain of length $L=12$, while all other dynamics are performed for $L=16$. The results for MBL dynamics are averaged over $1000$ disorder and initial state realizations, while all other dynamics are averaged over $200$ realizations.
    }
    \label{fig:HCEE_evolution}
\end{figure}

To complement these schematics, Fig.~\ref{fig:HCEE_evolution} displays the evolution of the half-chain entanglement entropy (HCEE) for each protocol, illustrating the trajectories toward their respective steady states. As shown, the HCEE for the NN thermal, NNN thermal, and mixed-field dynamics effectively saturates by $t=10^3$. In the MBL dynamics, the system reaches its characteristic steady state by $t=10^{12}$. Similarly, the RQC and Floquet dynamics achieve full saturation at a circuit depth of $2000$ and $100$ driving periods, respectively. The analysis throughout the main text utilizes these saturated data points to characterize the late-time entanglement landscape.

\section{The Level Spacing Ratios Distribution Analysis}\label{sec:sm_level_spacing}

In this section, we provide proofs for the thermal nature of various dynamics.
The distribution of adjacent level spacing ratios, $p(r)$, is a powerful probe for distinguishing between thermal and non-thermal phases, a method rooted in random matrix theory~\cite{PhysRevLett.52.1,
PhysRevB.75.155111,
PhysRevB.82.174411,
doi:10.1126/science.aao1401,
Mehta2004}. We apply this diagnostic to the four Hamiltonians ($\hat{H}_{\text{NN}}(W=0.5)$, $\hat{H}_{\text{NNN}}$, $\hat{H}_{\text{NN}}(W=5.0)$ and $\hat{H}_{\text{MF}}$) and the Floquet operator $\hat{F}$ to confirm their dynamical character.

We perform exact diagonalization for $\hat{H}_{\text{MF}}$ and other operators in the whole Hilbert space and within the half-filling sector where $\hat{S}^z_{\text{tot}}=\sum_{i=1}^{L}\hat{S}_{i}^z=0$, respectively. For the Hamiltonian systems, we analyze the middle one-third of the sorted eigenenergy spectrum, a standard procedure consistent with Ref.~\cite{PhysRevB.82.174411}. For Floquet operator $\hat{F}=\mathrm{e}^{-\mathrm{i}T_0\hat{H}_0}\mathrm{e}^{-\mathrm{i}T_1\hat{H}_1}$, we analyze all of its eigenvalues $\left\lbrace\mathrm{e}^{-\mathrm{i}\theta_k}\right\rbrace$ and set the quasienergies $\left\lbrace\theta_k\right\rbrace$ within range $[-\pi,\pi)$.

\begin{figure}[!htbp]
    \centering
    \includegraphics[width=1.0\columnwidth]{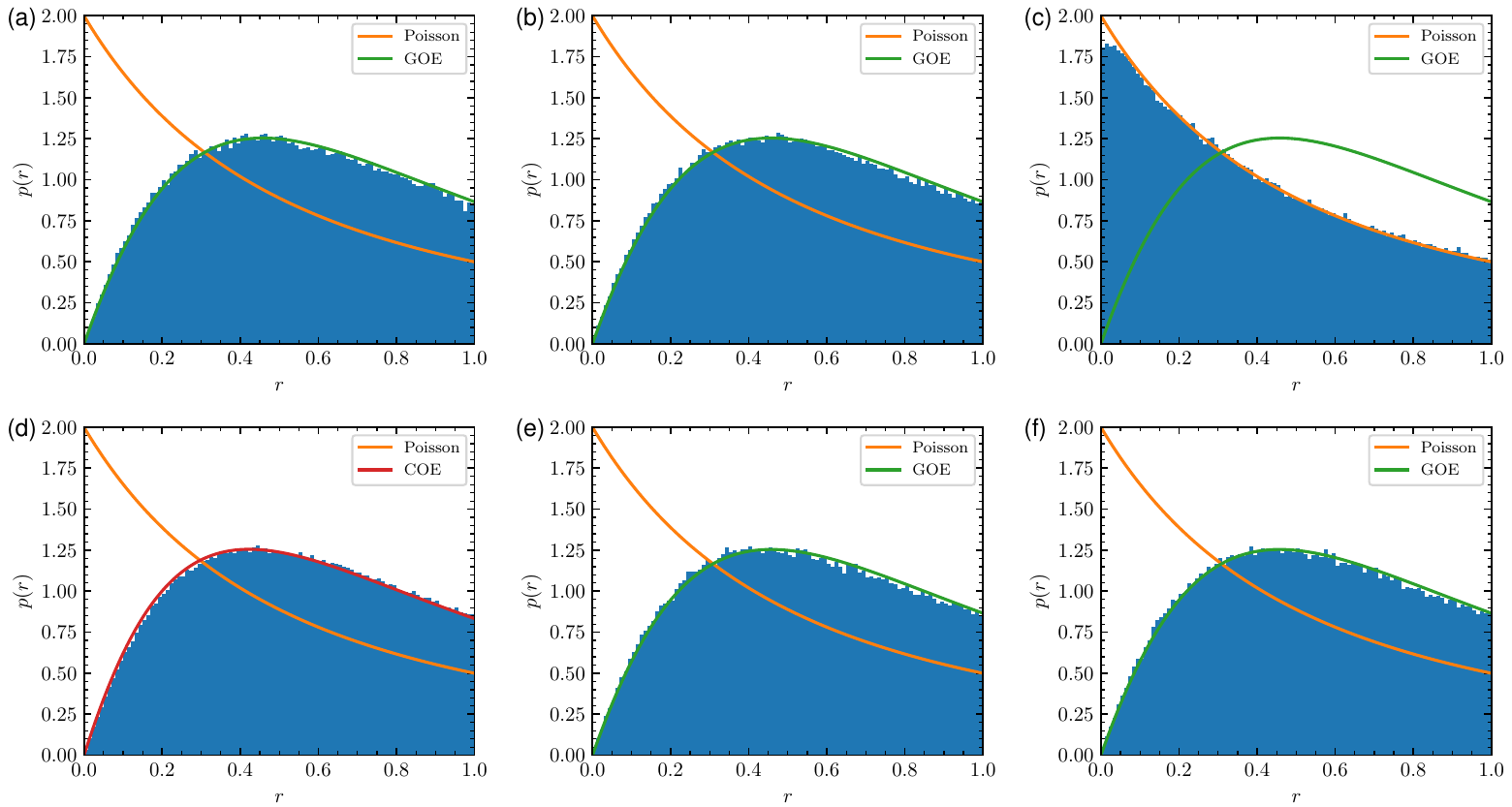}
    \caption{
    Level spacing ratio distributions $p(r)$ of various operators. For comparison, the theoretical predictions are overlaid as solid lines.
    Subplots (a-d) show the results within the half-filling sector of $\hat{H}_{\text{NN}}(W=0.5)$, $\hat{H}_{\text{NNN}}$, $\hat{H}_{\text{NN}}(W=5.0)$ and $\hat{F}$ for a system size of $L=16$. An additional result within the half-filling sector of $\hat{H}_{\text{NN}}(W=0.5)$ for a smaller system size of $L=12$ is shown in (f) as we consider this size for a dynamics governed by $\hat{H}_{\text{NN}}(W=0.5)$. Subplot (e) show the result in the whole Hilbert space of $\hat{H}_{\text{MF}}$ for $L=12$. Each histogram in (a-d) is compiled from $100$ independent samples with disorder realizations of corresponding strength and these numbers of samples in (e,f) are respectively $200$ and $1000$.
    (a) The NN thermal case for $\hat{H}_{\text{NN}}(W=0.5)$. $L=16$. $\bar{r}\approx0.531$.
    (b) The NNN thermal case for $\hat{H}_{\text{NNN}}$. $\bar{r}\approx0.530$.
    (c) The MBL case for $\hat{H}_{\text{NN}}(W=5.0)$. $\bar{r}\approx0.390$.
    (d) The Floquet thermal case for $\hat{F}=\mathrm{e}^{-\mathrm{i}T_0\hat{H}_0}\mathrm{e}^{-\mathrm{i}T_1\hat{H}_1}$. $\bar{r}\approx0.531$.
    (e) The mixed-field thermal case for $\hat{H}_{\text{MF}}$. $\bar{r}\approx0.531$.
    (f) The NN thermal case for $\hat{H}_{\text{NN}}(W=0.5)$. $L=12$. $\bar{r}\approx0.531$.
    }
    \label{fig:level_spacing}
\end{figure}

Given a list of eigenenergies $\left\lbrace E_k\right\rbrace$ or quasienergies $\left\lbrace\theta_k\right\rbrace$ sorted in ascending order, the consecutive level spacings are defined as $\delta_k = E_{k+1} - E_k$ or $\delta_k = \theta_{k+1} - \theta_k$, respectively. The level spacing ratio $r_k$ is then defined as the ratio of the smaller to the larger of two consecutive spacings:
\begin{equation}
r_k=\frac{\min\left(\delta_{k},\delta_{k+1}\right)}{\max\left(\delta_{k},\delta_{k+1}\right)}.
\label{eq:level_spacing_ratio}
\end{equation}
From the collection of all such ratios, we construct the histograms of $\left\lbrace r_k\right\rbrace$ of all cases, presented in Fig.~\ref{fig:level_spacing}, along with the numerical average values $\bar{r}$.

This diagnostic tool provides a sharp distinction between different dynamical regimes. For the thermal cases of our models, the histograms of $\left\lbrace r_k\right\rbrace$ for Hamiltonian systems and Floquet system are expected to follow the Gaussian Orthogonal Ensemble (GOE) and the Circular Orthogonal Ensemble (COE), respectively. In contrast, the MBL phase is characterized by a Poisson distribution. As shown in Fig.~\ref{fig:level_spacing}, our numerical results align perfectly with corresponding theoretical baselines. The distributions for the three thermal Hamiltonians and the Floquet operator match the GOE and COE curves, confirming they are in the thermal regime. Conversely, the MBL Hamiltonian follows the Poisson distribution, a clear signature of its MBL nature. All these are further corroborated by $\bar{r}$ for each case, which closely matches the theoretical average value of the corresponding distribution: $\langle r\rangle_\text{GOE}=4-2\sqrt3\approx0.536$ \cite{PhysRevLett.110.084101}, $\langle r\rangle_\text{COE}\approx0.527$ \cite{PhysRevX.4.041048} and $\langle r\rangle_\text{Poisson}=2\ln2-1\approx0.386$ \cite{PhysRevLett.110.084101}.

\section{Symmetry Analysis and Enumeration of Bipartition Geometries}\label{sec:sm_bipartitions_symmetry_analysis}

A central aspect of our methodology is the simplification of the analysis by exploiting the system's symmetries. As mentioned in the main text, our analysis involves averaging over various sources of randomness (e.g., initial states, disorders, and circuit realizations). This procedure, combined with the PBC, restores both translational and parity symmetries to the ensemble-averaged system.

These symmetries have two crucial consequences that greatly reduce the complexity of our analysis:
\begin{enumerate}
    \item The mutual information between any two spins depends solely on their separation distance $j$, and we denote it as $I_j$.
    \item The analysis of EE across all $2^{L-1}-1$ nontrivial bipartitions can be reduced to studying a much smaller, minimal set of symmetry-inequivalent representative bipartitions.
\end{enumerate}

For a one-dimensional spin chain of even length $L$, there are $2^L$ ways to form a subsystem $\text{A}$. However, since the EE is symmetric with respect to the subsystem and its complement, we only consider bipartitions where the subsystem size $n_0$ does not exceed half the chain length, i.e., $n_0\leqslant L/2$. The case $n_0=0$ is trivial. Thus, the total number of distinct, nontrivial bipartitions is given by $\sum_{n_0=1}^{L/2-1}\binom{L}{n_0}+\frac{1}{2}\binom{L}{L/2}=2^{L-1}-1$. Bipartitions that can be mapped onto one another via translation or parity operation are considered symmetry-equivalent, as they yield identical EE values upon averaging. We therefore only need to consider a reduced set of representative bipartitions. We denote the number of such representative bipartitions with chain length $L$ and a fixed subsystem size $n_0$ as $N(L,n_0)$.

Our multi-bipartition entanglement tomography is based on the geometric characterization of a bipartition via its crossed-bond vector $\left(n_1,n_2,\cdots,n_{L/2-1}\right)$. While symmetry-equivalent bipartitions are guaranteed to have identical crossed-bond vectors, having the same crossed-bond vector does not necessarily imply that two bipartitions are symmetry-equivalent. In fact, two bipartitions that are inequivalent under symmetry operations can accidentally possess the same crossed-bond vector.

Notably, our numerical results indicate that such symmetry-inequivalent bipartitions, even when sharing an identical crossed-bond vector, can yield slightly different average EE values. This nontrivial difference means that each of the $N\left(L,n_0\right)$ representative bipartitions constitutes a distinct data point $\left(n_1,n_2,\cdots,n_{L/2-1},S\right)$ for our fitting framework. Therefore, for each fixed subsystem size $n_0$ with chain length $L$, the fit is performed on a dataset of $N\left(L,n_0\right)$ samples.

\begin{table}[!htbp]
\caption{For a spin chain with PBC and length $L=12$ or $16$, this table lists the number of symmetry-inequivalent representative bipartitions, $N\left(L,n_0\right)$, and the number of unique crossed-bond geometries, $M\left(L,n_0\right)$, for each subsystem size $n_0$.}
\begin{ruledtabular}
\begin{tabular}{ccccccccc}
$n_0$&
1&
2&
3&
4&
5&
6&
7&
8\\ \hline
$N\left(16,n_0\right)$&
1&
8&
21&
72&
147&
280&
375&
257
\\
$M\left(16,n_0\right)$&
1&
8&
21&
70&
137&
246&
327&
254
\\
$N\left(12,n_0\right)$&
1&
6&
12&
29&
38&
35&
 &
 
\\
$M\left(12,n_0\right)$&
1&
6&
12&
28&
35&
35&
 &
 
\\
\end{tabular}
\end{ruledtabular}
\label{tab:n_0_N_M}
\end{table}

To assess the diversity of our independent variables, we also count the number of \textit{unique} crossed-bond vectors with fixed subsystem size, denoted by $M\left(L,n_0\right)$. This quantity represents the number of distinct locations in the predictor space of our regression. By definition, $M\left(L,n_0\right)\leqslant N\left(L,n_0\right)$. As detailed in Table~\ref{tab:n_0_N_M}, the values of $M\left(L,n_0\right)$ are only marginally smaller than those of $N\left(L,n_0\right)$ for the cases we considered in this paper. This confirms that the geometric configurations sampled are not highly degenerate and are well-distributed throughout the predictor space (in fact, for the situations listed in Table~\ref{tab:n_0_N_M}, the maximum degeneracy does not exceed $3$), thus ensuring the reliability of our fitting procedure.

\section{More Results about Mutual Information}\label{sec:sm_mutual_information}

\begin{figure}[!htbp]
    \centering
    \includegraphics[width=0.4\columnwidth]{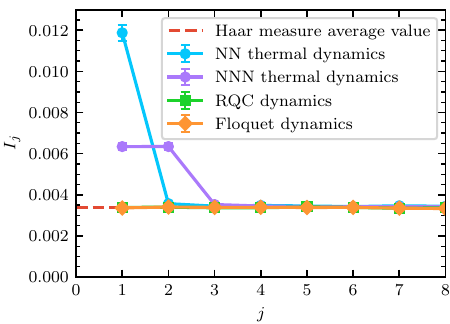}
    \caption{
    Saturated mutual informations $\left\lbrace I_j\right\rbrace$ of various thermal dynamics for $L=16$. Results are shown at evolution time $t=1000.0$ for NN and NNN thermal dynamics, a circuit depth of 2000 for RQC, and after 100 periods for Floquet dynamics. All data are averaged over 1000 random samples.
    }
    \label{fig:thermal_sat_I}
\end{figure}

To contextualize the hierarchy of entanglement bond tensions $\{\omega_j\}$ discussed in the main text, this section provides a systematic comparison of the saturated mutual information $\{I_j\}$ across various thermal dynamics (Fig.~\ref{fig:thermal_sat_I}). As observed with multi-bipartition entanglement tomography, the RQC and Floquet dynamics perfectly reproduce the Haar measure average. In contrast, the two Hamiltonian dynamics exhibit short-range mutual information, specifically at distances where direct spin interactions occur, that is significantly higher than at longer distances. These trends in mutual information closely mirror the results obtained for the entanglement bond tensions across all dynamical protocols. 

\section{Detailed Analysis of Fitting Quality and More Fitting Results}\label{sec:sm_fitting_details}

This section provides a systematic analysis of the linear fitting results derived from our multi-bipartition entanglement tomography.  To demonstrate the generality of our findings, we present results not only for the equal-sized bipartitions ($n_0 = L/2$) discussed in the main text, but also for several cases of unequal bipartitions ($n_0 < L/2$). We focus on relatively large values of $n_0$ as they involve a richer set of crossed-bond geometries and have smaller finite size effect. For each dynamical protocol, we select representative time points (or evolution depths/periods) to illustrate the entanglement bond tensions $\left\lbrace \omega_j\right\rbrace$ at different stages of the evolution. The following figures visualize the results in different protocols, offering clear evidence for the robustness of the bond entanglement tension hierarchy and the bond-additive law discussed in the main text.

\begin{figure}[!htbp]
    \centering
    \includegraphics[width=0.85\columnwidth]{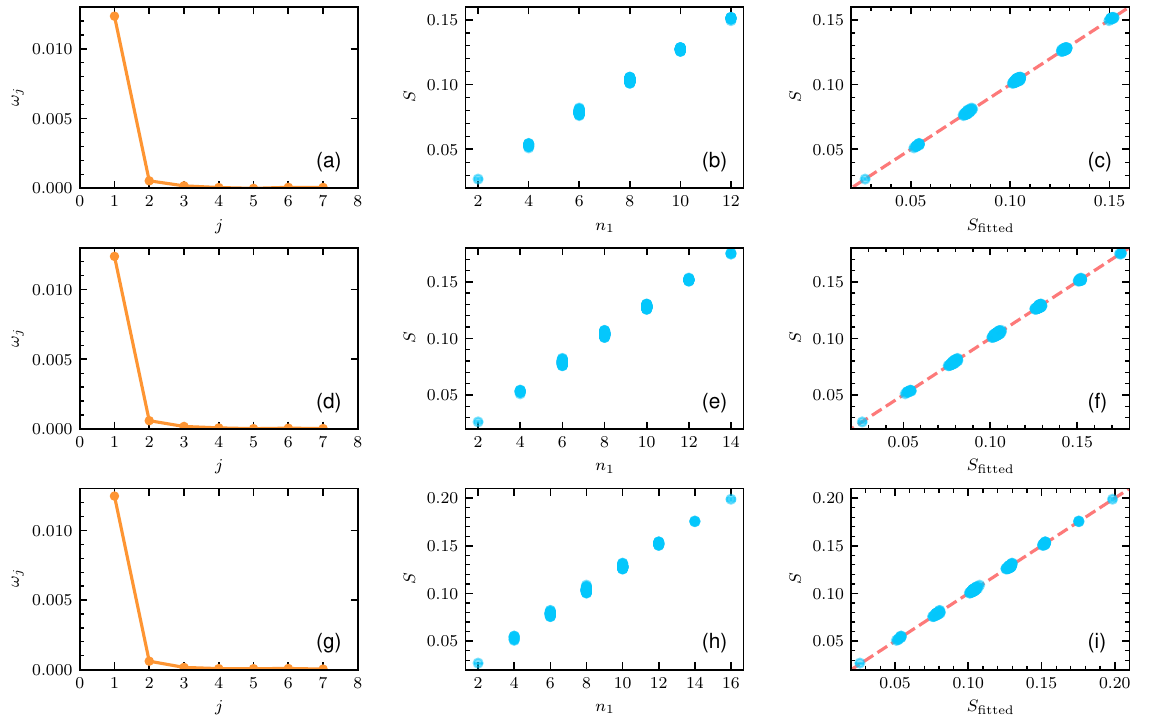}
    \caption{
    Results of NN thermal dynamics for $L=16$. Evolution time $t=0.1$. (a-c) $n_0=6$, $R^2\approx0.9994$. (d-f) $n_0=7$, $R^2\approx0.9993$. (g-i) $n_0=8$, $R^2\approx0.9993$.
    }
    \label{fig:NN_thermal_fig_t=0.1}
\end{figure}

\begin{figure}[!htbp]
    \centering
    \includegraphics[width=0.85\columnwidth]{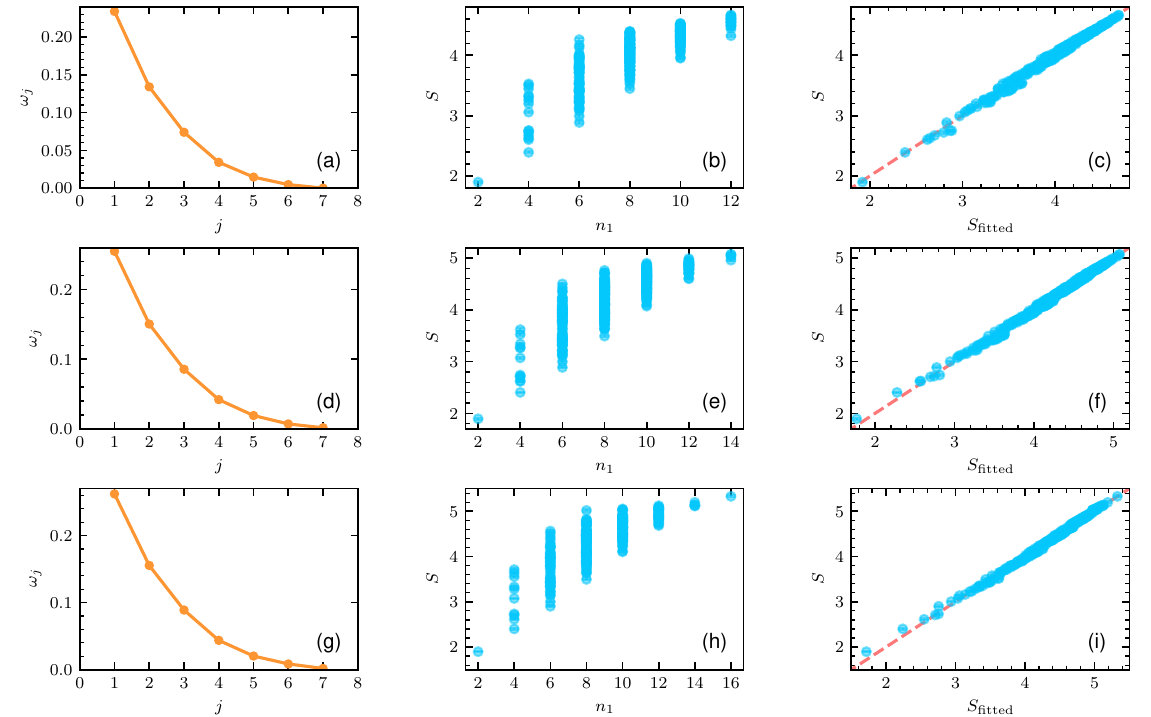}
    \caption{
    Results of NN thermal dynamics for $L=16$. Evolution time $t=2.0$. (a-c) $n_0=6$, $R^2\approx0.9933$. (d-f) $n_0=7$, $R^2\approx0.9957$. (g-i) $n_0=8$, $R^2\approx0.9959$.
    }
    \label{fig:NN_thermal_fig_t=2.0}
\end{figure}

\begin{figure}[!htbp]
    \centering
    \includegraphics[width=0.85\columnwidth]{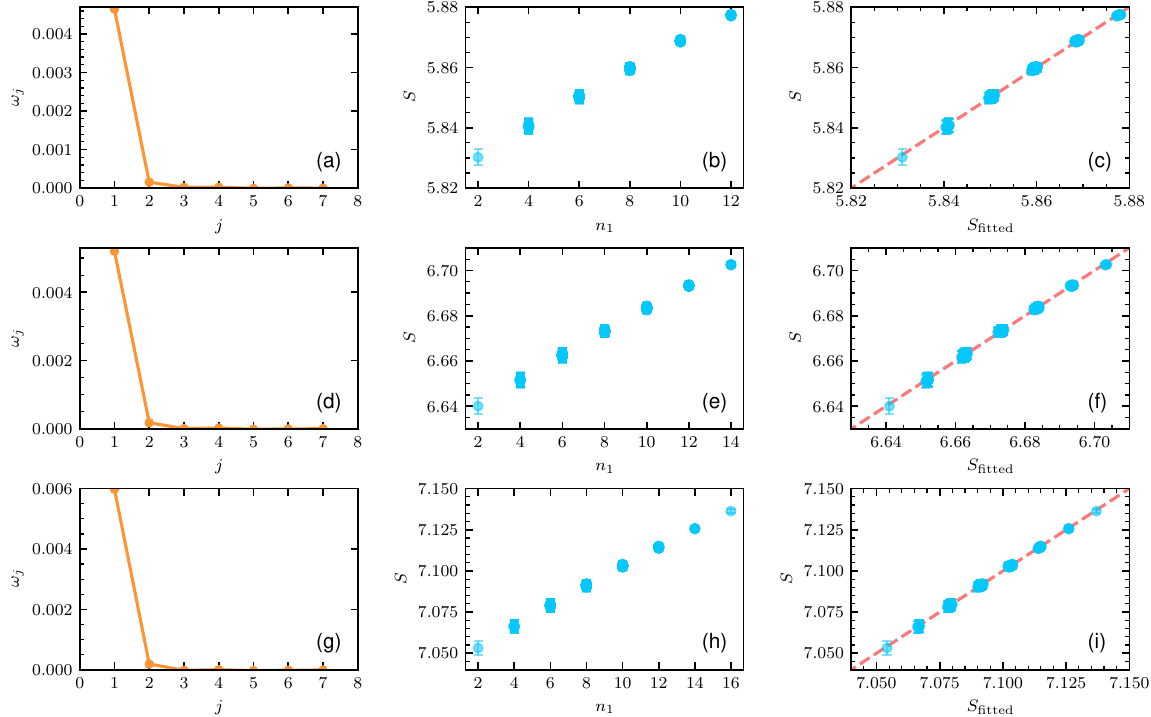}
    \caption{
    Results of NN thermal dynamics for $L=16$. Evolution time $t=1000.0$. (a-c) $n_0=6$, $R^2\approx0.9994$. (d-f) $n_0=7$, $R^2\approx0.9994$. (g-i) $n_0=8$, $R^2\approx0.9995$.
    }
    \label{fig:NN_thermal_fig_t=1000.0}
\end{figure}

\begin{figure}[!htbp]
    \centering
    \includegraphics[width=0.85\columnwidth]{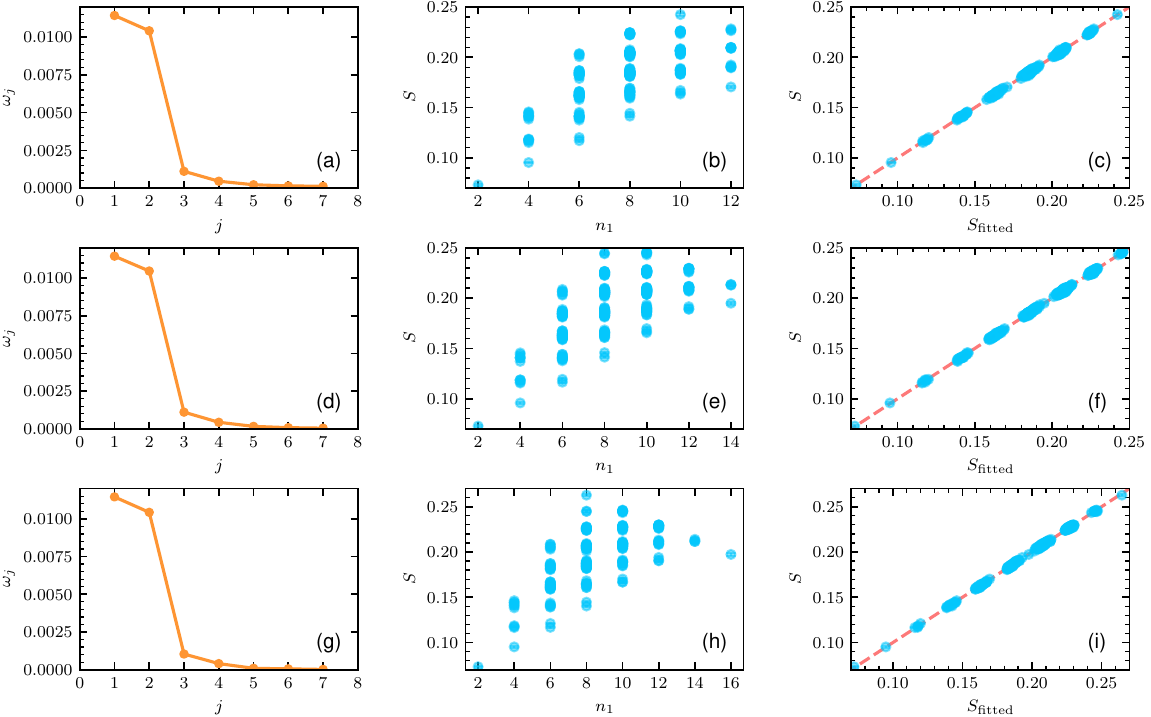}
    \caption{
    Results of NNN thermal dynamics for $L=16$. Evolution time $t=0.1$. (a-c) $n_0=6$, $R^2\approx0.9989$. (d-f) $n_0=7$, $R^2\approx0.9991$. (g-i) $n_0=8$, $R^2\approx0.9992$.
    }
    \label{fig:NNN_thermal_fig_t=0.1}
\end{figure}

\begin{figure}[!htbp]
    \centering
    \includegraphics[width=0.85\columnwidth]{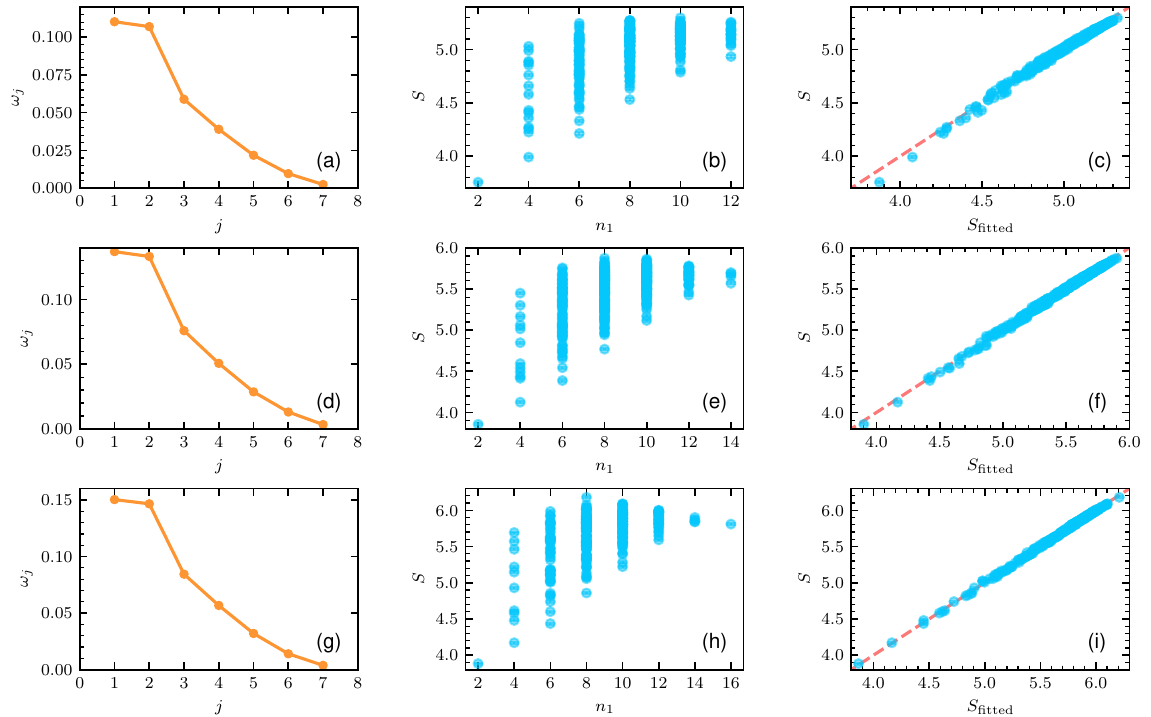}
    \caption{
    Results of NNN thermal dynamics for $L=16$. Evolution time $t=2.0$. (a-c) $n_0=6$, $R^2\approx0.9923$. (d-f) $n_0=7$, $R^2\approx0.9974$. (g-i) $n_0=8$, $R^2\approx0.9989$.
    }
    \label{fig:NNN_thermal_fig_t=2.0}
\end{figure}

\begin{figure}[!htbp]
    \centering
    \includegraphics[width=0.85\columnwidth]{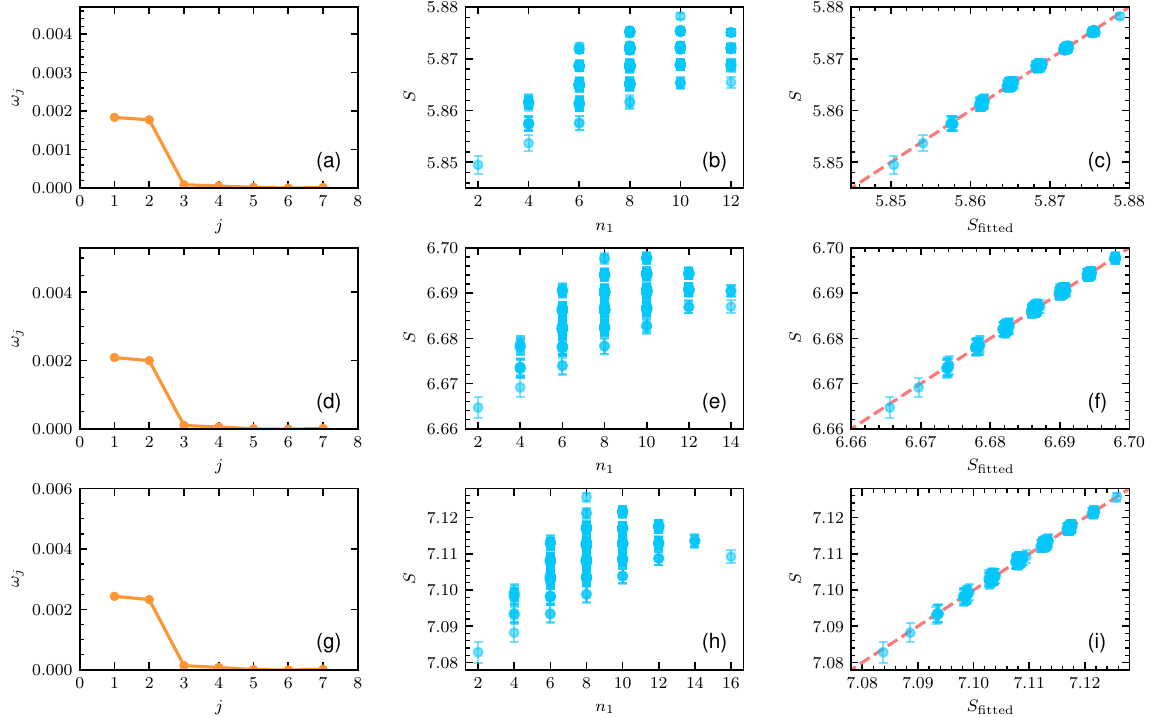}
    \caption{
    Results of NNN thermal dynamics for $L=16$. Evolution time $t=1000.0$. (a-c) $n_0=6$, $R^2\approx0.9979$. (d-f) $n_0=7$, $R^2\approx0.9982$. (g-i) $n_0=8$, $R^2\approx0.9986$.
    }
    \label{fig:NNN_thermal_fig_t=1000.0}
\end{figure}

\begin{figure}[!htbp]
    \centering
    \includegraphics[width=0.85\columnwidth]{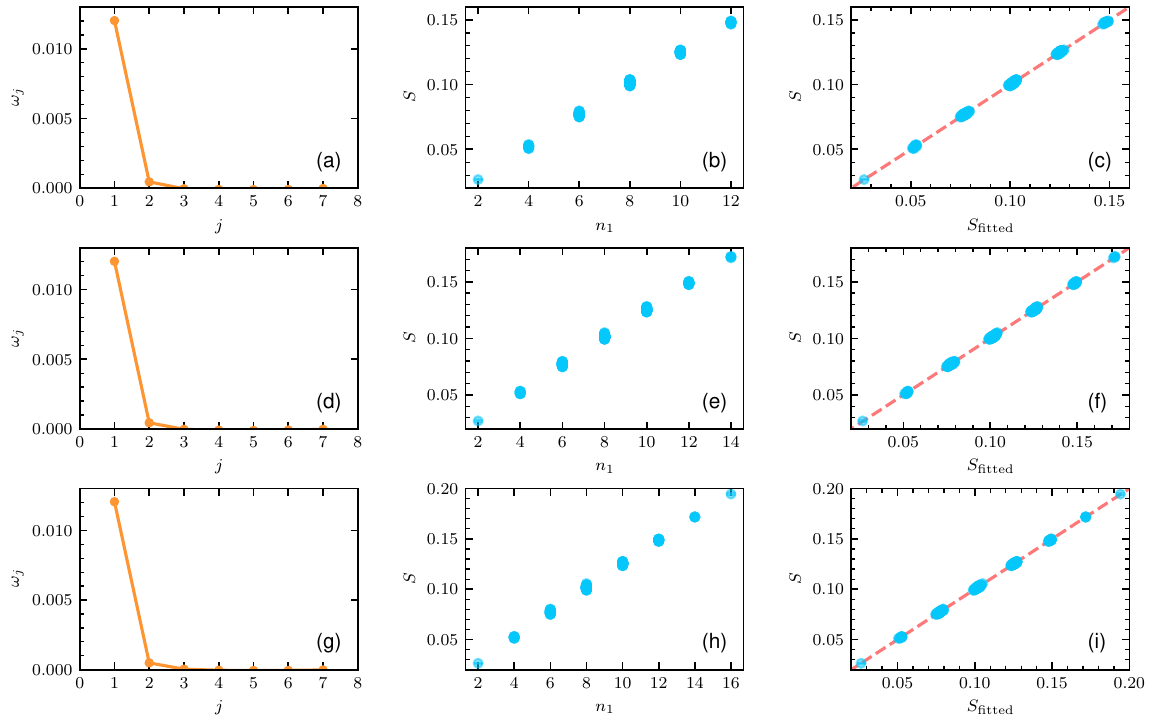}
    \caption{
    Results of MBL dynamics for $L=16$. Evolution time $t=0.1$. (a-c) $n_0=6$, $R^2\approx0.9996$. (d-f) $n_0=7$, $R^2\approx0.9997$. (g-i) $n_0=8$, $R^2\approx0.9998$.
    }
    \label{fig:MBL_fig_t=0.1}
\end{figure}

\begin{figure}[!htbp]
    \centering
    \includegraphics[width=0.85\columnwidth]{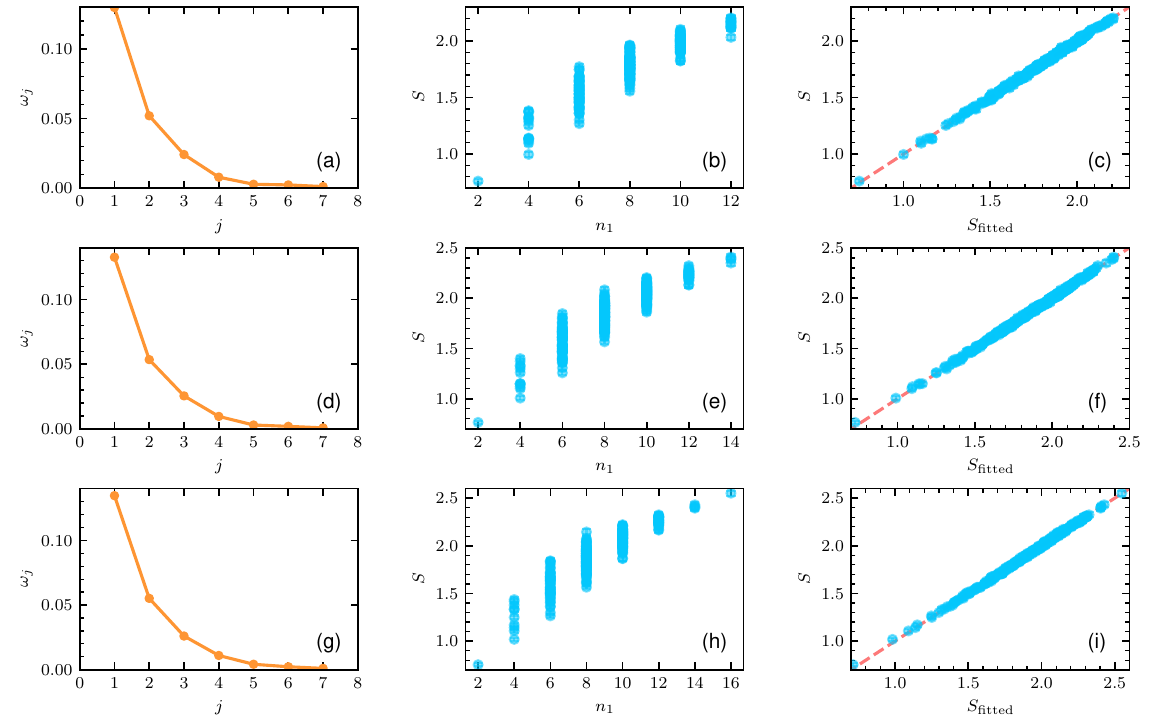}
    \caption{
    Results of MBL dynamics for $L=16$. Evolution time $t=10.0$. (a-c) $n_0=6$, $R^2\approx0.9976$. (d-f) $n_0=7$, $R^2\approx0.9981$. (g-i) $n_0=8$, $R^2\approx0.9984$.
    }
    \label{fig:MBL_fig_t=10.0}
\end{figure}

\begin{figure}[!htbp]
    \centering
    \includegraphics[width=0.85\columnwidth]{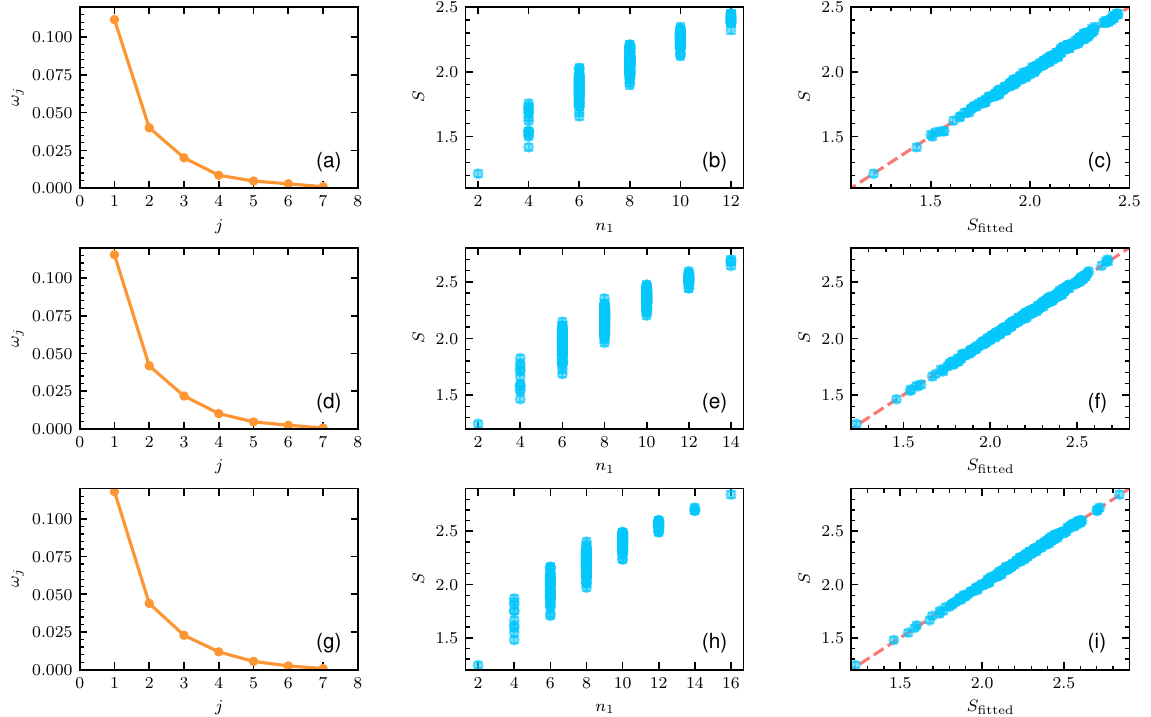}
    \caption{
    Results of MBL dynamics for $L=16$. Evolution time $t=1000.0$. (a-c) $n_0=6$, $R^2\approx0.9972$. (d-f) $n_0=7$, $R^2\approx0.9976$. (g-i) $n_0=8$, $R^2\approx0.9981$.
    }
    \label{fig:MBL_fig_t=1000.0}
\end{figure}

\begin{figure}[!htbp]
    \centering
    \includegraphics[width=0.85\columnwidth]{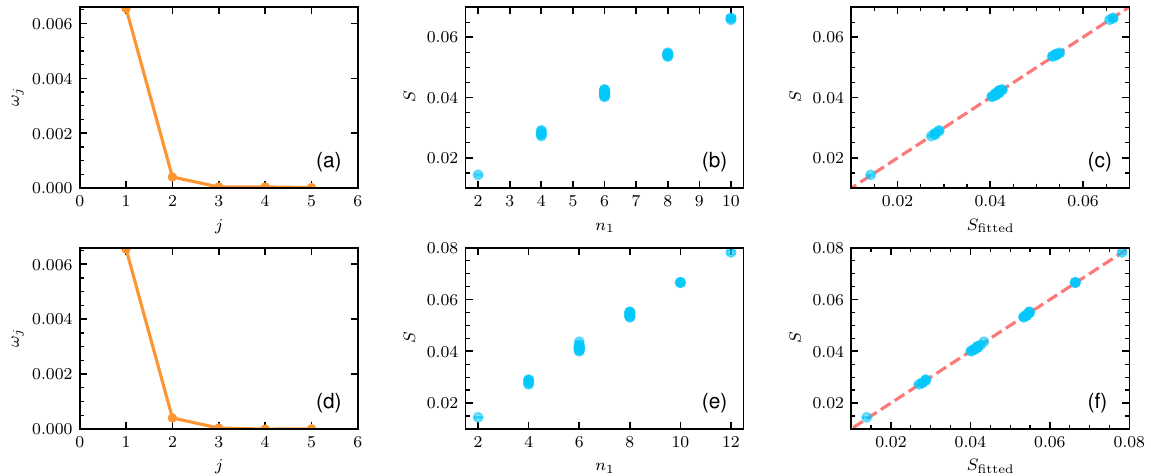}
    \caption{
    Results of mixed-field dynamics for $L=12$. Evolution time $t=0.1$. (a-c) $n_0=5$, $R^2\approx0.9997$. (d-f) $n_0=6$, $R^2\approx0.9997$.
    }
    \label{fig:MF_fig_t=0.1}
\end{figure}

\begin{figure}[!htbp]
    \centering
    \includegraphics[width=0.85\columnwidth]{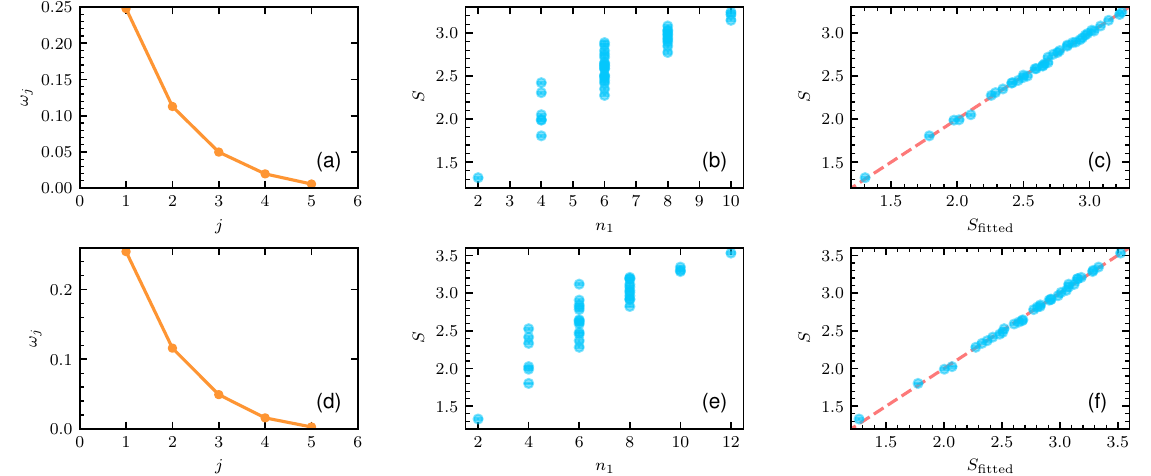}
    \caption{
    Results of mixed-field dynamics for $L=12$. Evolution time $t=2.0$. (a-c) $n_0=5$, $R^2\approx0.9978$. (d-f) $n_0=6$, $R^2\approx0.9975$.
    }
    \label{fig:MF_fig_t=2.0}
\end{figure}

\begin{figure}[!htbp]
    \centering
    \includegraphics[width=0.85\columnwidth]{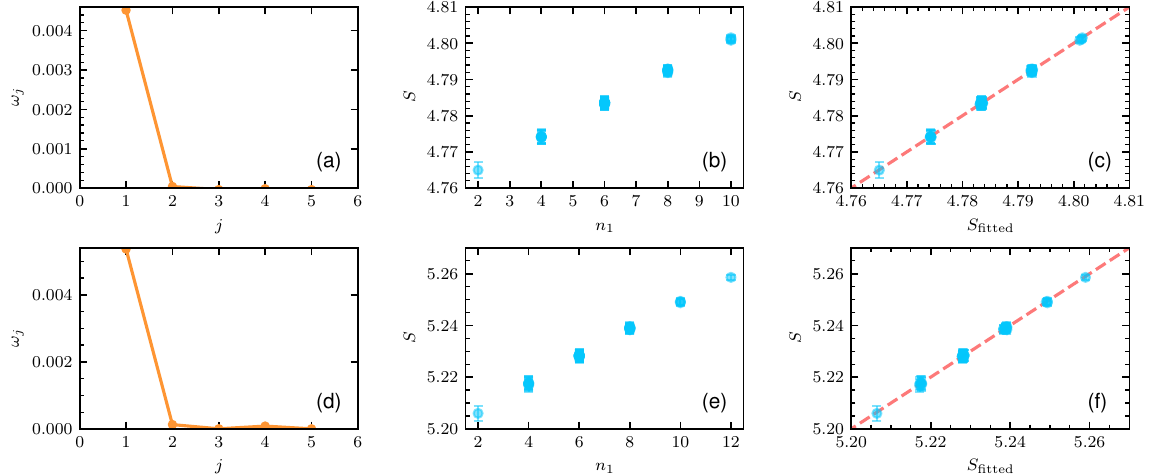}
    \caption{
    Results of mixed-field dynamics for $L=12$. Evolution time $t=1000.0$. (a-c) $n_0=5$, $R^2\approx0.9991$. (d-f) $n_0=6$, $R^2\approx0.9990$.
    }
    \label{fig:MF_fig_t=1000.0}
\end{figure}

\begin{figure}[!htbp]
    \centering
    \includegraphics[width=0.85\columnwidth]{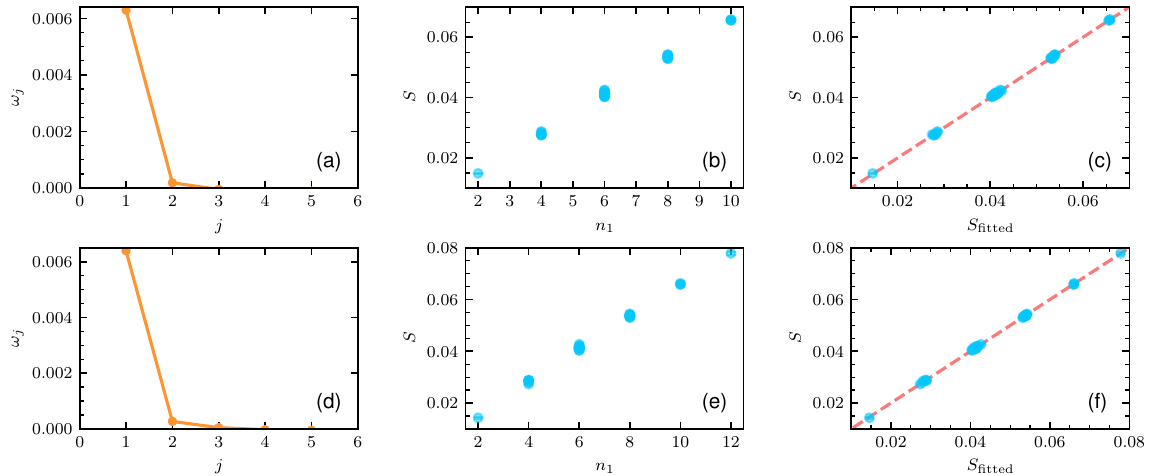}
    \caption{
    Results of dynamics governed by $\hat{H}_{\mathrm{NN}}(W=0.5)$ starting from random product state for $L=12$. Evolution time $t=0.1$. (a-c) $n_0=5$, $R^2\approx0.9996$. (d-f) $n_0=6$, $R^2\approx0.9998$.
    }
    \label{fig:UPN_fig_t=0.1}
\end{figure}

\begin{figure}[!htbp]
    \centering
    \includegraphics[width=0.85\columnwidth]{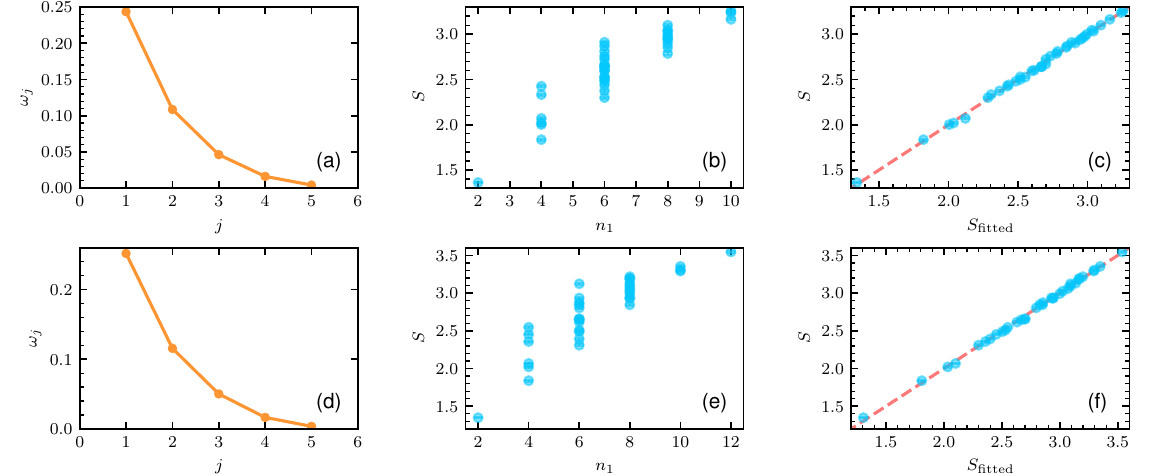}
    \caption{
    Results of dynamics governed by $\hat{H}_{\mathrm{NN}}(W=0.5)$ starting from random product state for $L=12$. Evolution time $t=2.0$. (a-c) $n_0=5$, $R^2\approx0.9980$. (d-f) $n_0=6$, $R^2\approx0.9983$.
    }
    \label{fig:UPN_fig_t=2.0}
\end{figure}

\begin{figure}[!htbp]
    \centering
    \includegraphics[width=0.85\columnwidth]{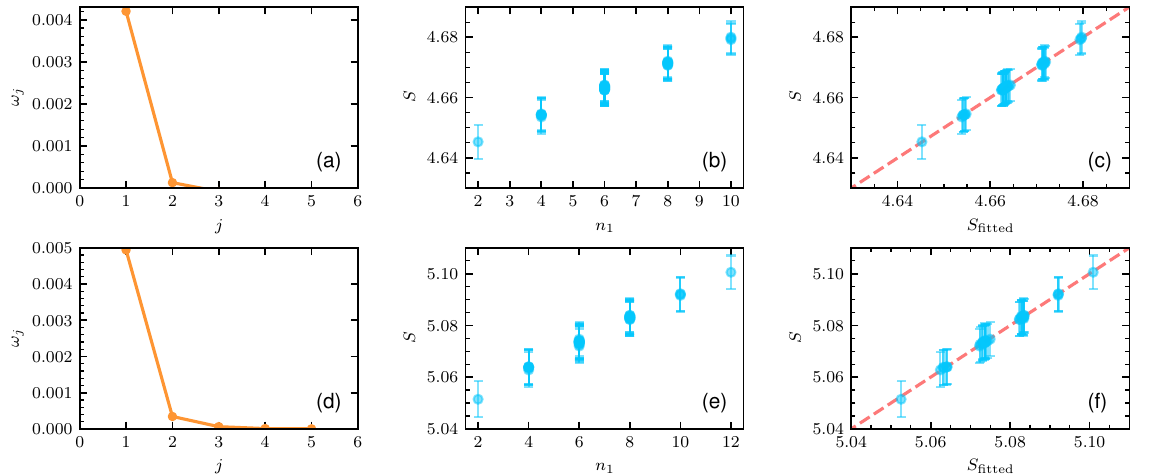}
    \caption{
    Results of dynamics governed by $\hat{H}_{\mathrm{NN}}(W=0.5)$ starting from random product state for $L=12$. Evolution time $t=1000.0$. (a-c) $n_0=5$, $R^2\approx0.9991$. (d-f) $n_0=6$, $R^2\approx0.9986$.
    }
    \label{fig:UPN_fig_t=1000.0}
\end{figure}

\begin{figure}[!htbp]
    \centering
    \includegraphics[width=0.85\columnwidth]{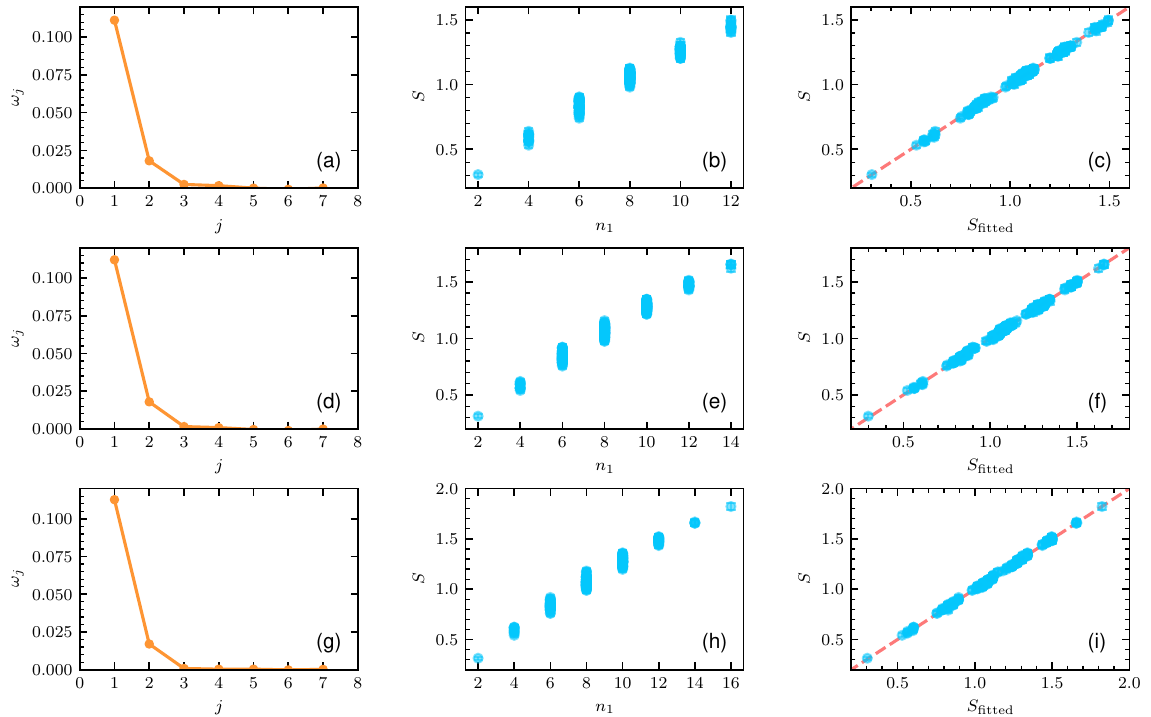}
    \caption{
    Results of RQC dynamics for $L=16$. Evolution depth is $5$. (a-c) $n_0=6$, $R^2\approx0.9975$. (d-f) $n_0=7$, $R^2\approx0.9975$. (g-i) $n_0=8$, $R^2\approx0.9978$.
    }
    \label{fig:RQC_fig_depth=5}
\end{figure}

\begin{figure}[!htbp]
    \centering
    \includegraphics[width=0.85\columnwidth]{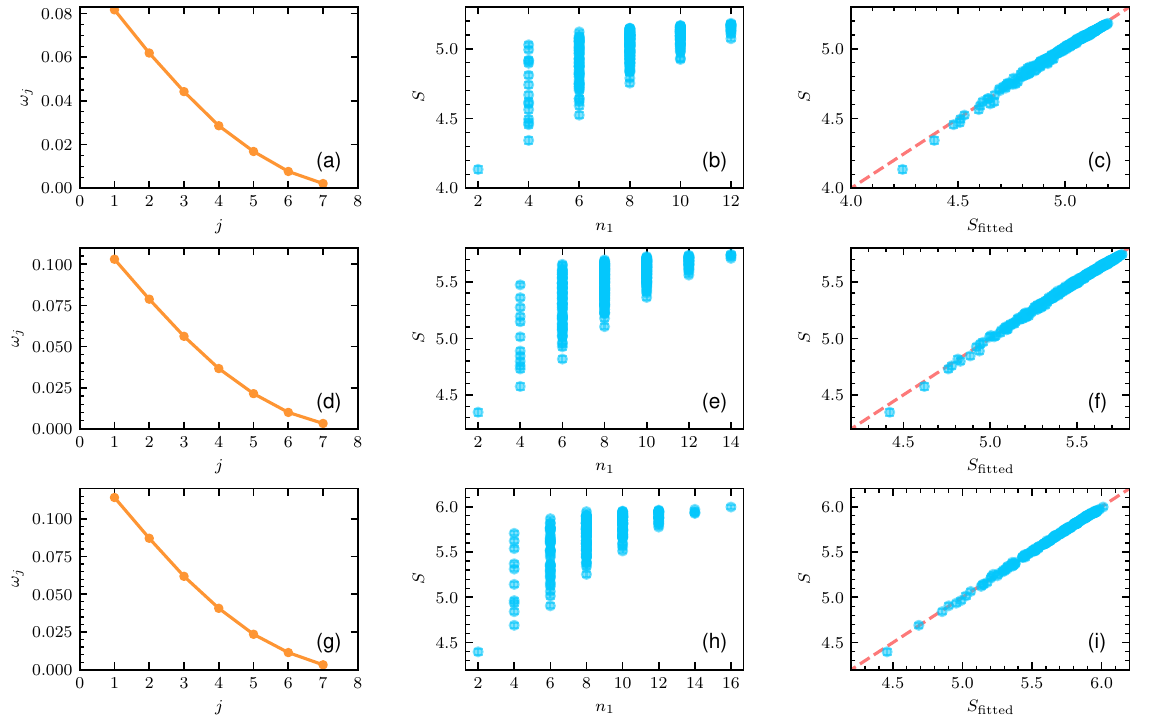}
    \caption{
    Results of RQC dynamics for $L=16$. Evolution depth is $100$. (a-c) $n_0=6$, $R^2\approx0.9919$. (d-f) $n_0=7$, $R^2\approx0.9968$. (g-i) $n_0=8$, $R^2\approx0.9986$.
    }
    \label{fig:RQC_fig_depth=100}
\end{figure}

\begin{figure}[!htbp]
    \centering
    \includegraphics[width=0.85\columnwidth]{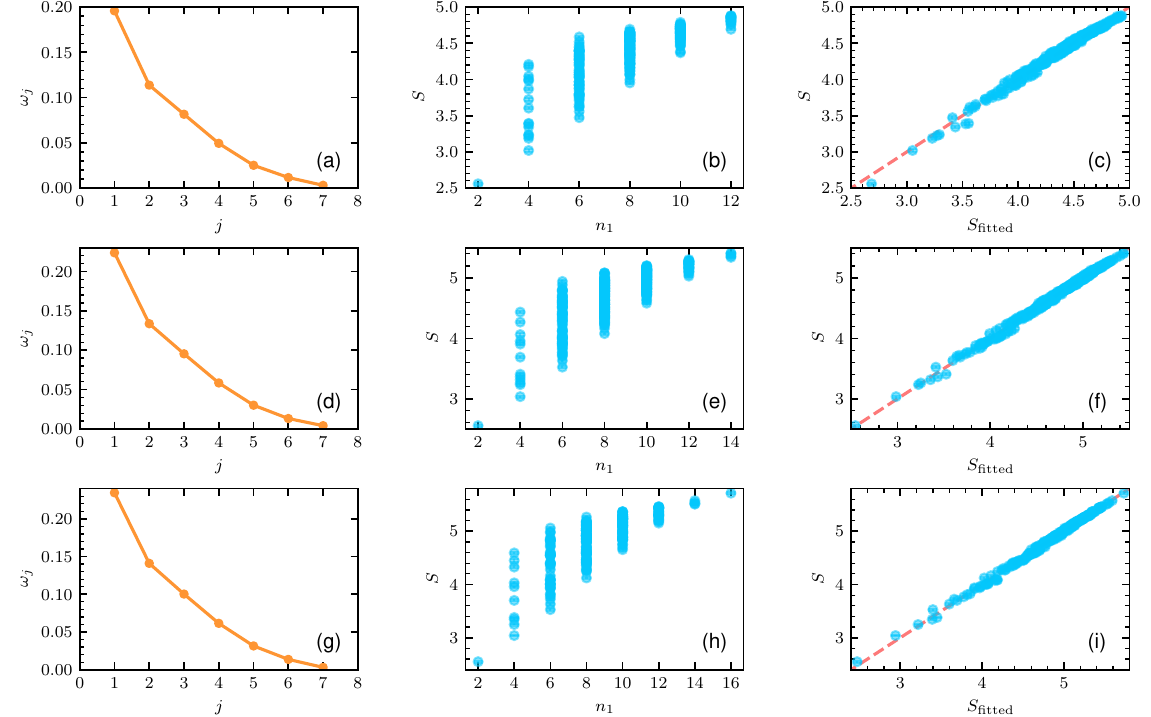}
    \caption{
    Results of Floquet dynamics for $L=16$. The period number of evolution is $1$. (a-c) $n_0=6$, $R^2\approx0.9899$. (d-f) $n_0=7$, $R^2\approx0.9955$. (g-i) $n_0=8$, $R^2\approx0.9964$.
    }
    \label{fig:Floquet_fig_num_periods=1}
\end{figure}

\begin{figure}[!htbp]
    \centering
    \includegraphics[width=0.85\columnwidth]{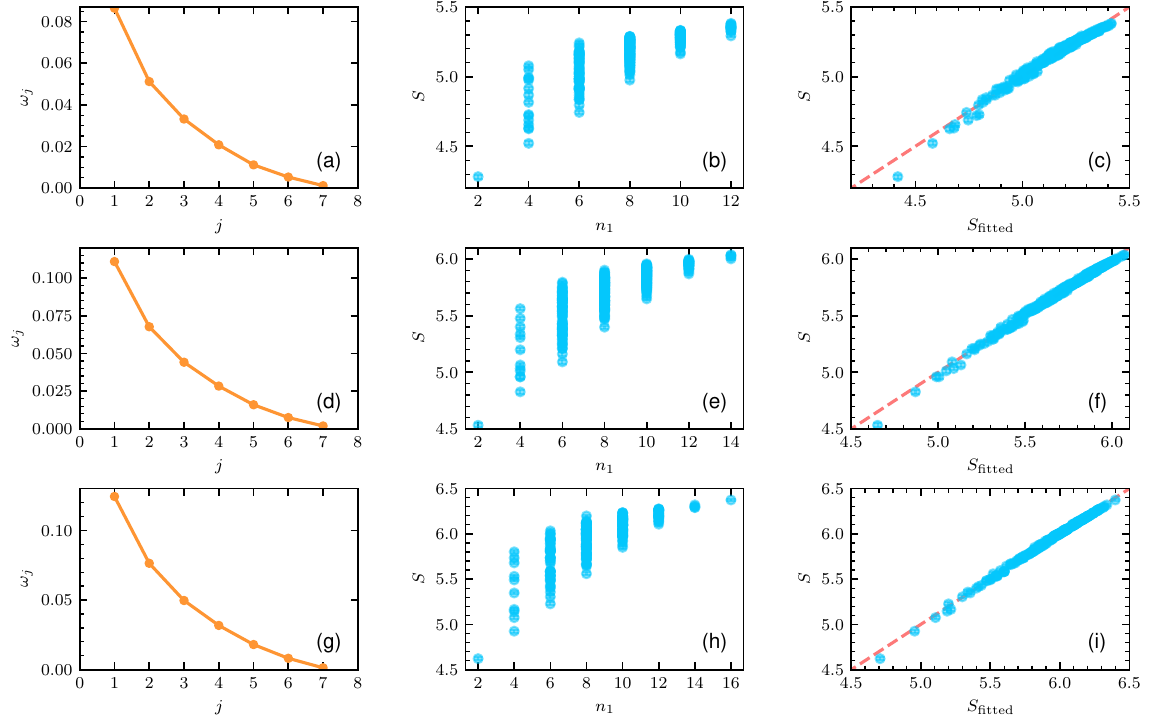}
    \caption{
    Results of Floquet dynamics for $L=16$. The period number of evolution is $3$. (a-c) $n_0=6$, $R^2\approx0.9836$. (d-f) $n_0=7$, $R^2\approx0.9942$. (g-i) $n_0=8$, $R^2\approx0.9977$.
    }
    \label{fig:Floquet_fig_num_periods=3}
\end{figure}

All the data of EE used for fitting are averaged over $1000$ random samples. Each figure is organized into a three-column format to offer a comprehensive view of the fitting quality and the underlying physical data: The first column shows the extracted entanglement bond tensions $\left\lbrace \omega_j\right\rbrace$ as a function of the bond order $j$. This reveals the geometric structure of entanglement. The second column shows raw distribution of the EE $S$ as a function of the number of NN crossed bonds, $n_1$. This visualizes the raw data's structure. The third column shows a direct comparison between the actual measured EE values ($S$) and the values predicted by our bond-additive model ($S_{\text{fitted}}$). The dashed red line represents the ideal case $S = S_{\text{fitted}}$. The tight clustering of data points around this line visually confirms the extremely high quality and predictive power of our multi-bipartition entanglement tomography. Fig.~\ref{fig:NN_thermal_fig_t=0.1} to Fig.~\ref{fig:NN_thermal_fig_t=1000.0} are the case of NN thermal dynamics. Fig.~\ref{fig:NNN_thermal_fig_t=0.1} to Fig.~\ref{fig:NNN_thermal_fig_t=1000.0} are the case of NNN thermal dynamics. Fig.~\ref{fig:MBL_fig_t=0.1} to Fig.~\ref{fig:MBL_fig_t=1000.0} are the case of MBL dynamics. Fig.~\ref{fig:MF_fig_t=0.1} to Fig.~\ref{fig:MF_fig_t=1000.0} are the case of mixed-field dynamics. Fig.~\ref{fig:UPN_fig_t=0.1} to Fig.~\ref{fig:UPN_fig_t=1000.0} are the case of dynamics governed by $\hat{H}_{\mathrm{NN}}(W=0.5)$ starting from random product state without charge $U(1)$ conservation, defined in \Eq{random_product_state}. Fig.~\ref{fig:RQC_fig_depth=5} and Fig.~\ref{fig:RQC_fig_depth=100} are the case of RQC dynamics. Fig.~\ref{fig:Floquet_fig_num_periods=1} and Fig.~\ref{fig:Floquet_fig_num_periods=3} are the case of Floquet dynamics.








\end{document}